\documentclass[12pt,preprint]{aastex}
\input epsf
\usepackage{epsfig}

\begin{document}

\title{Astrophysical Neutrino Event Rates and Sensitivity for Neutrino
Telescopes}

\author{Ivone F.\ M. Albuquerque}
\affil{Department of Astronomy and Space Sciences Laboratory,
University of California, Berkeley, CA 94720.}
\email{IFAlbuquerque@lbl.gov}

\author{Jodi Lamoureux}
\affil{National Energy Research Scientific Computing Center, Lawrence Berkeley National Laboratory, Berkeley, CA 94720.} 
\email{JILamoureux@lbl.gov}

\and

\author{George F. Smoot}
\affil{Lawrence Berkeley National Laboratory, Space Sciences Laboratory and
Department of Physics, University of California, Berkeley, CA 94720.}
\email{GFSmoot@lbl.gov}

\begin{abstract}
Spectacular processes in astrophysical sites produce
high-energy cosmic rays which are further accelerated by Fermi-shocks
into a power-law spectrum. 
These, in passing through radiation fields and matter, produce
neutrinos.
Neutrino telescopes are designed with large detection volumes 
to observe such astrophysical sources.
A large volume is necessary because the fluxes and cross-sections
are small.
We estimate various telescopes' sensitivities 
and expected event rates from astrophysical 
sources of high-energy neutrinos.
We find that an ideal detector of km$^2$ incident area 
can be sensitive to a flux of neutrinos integrated over
energy from $10^{5}$ and $10^{7}$ GeV 
as low as $1.3 \times 10^{-8} \rm{ E}^{-2}$ (GeV/cm$^2$ s sr)
which is three times smaller than the Waxman-Bachall conservative
upper limit on potential neutrino flux.  
A real detector will have degraded performance. 
Detection from known
point sources is possible but unlikely unless there is 
prior knowledge of the source location and neutrino arrival time.
\end{abstract}


\newpage

\section{Introduction}

\begin{figure}
\plotone{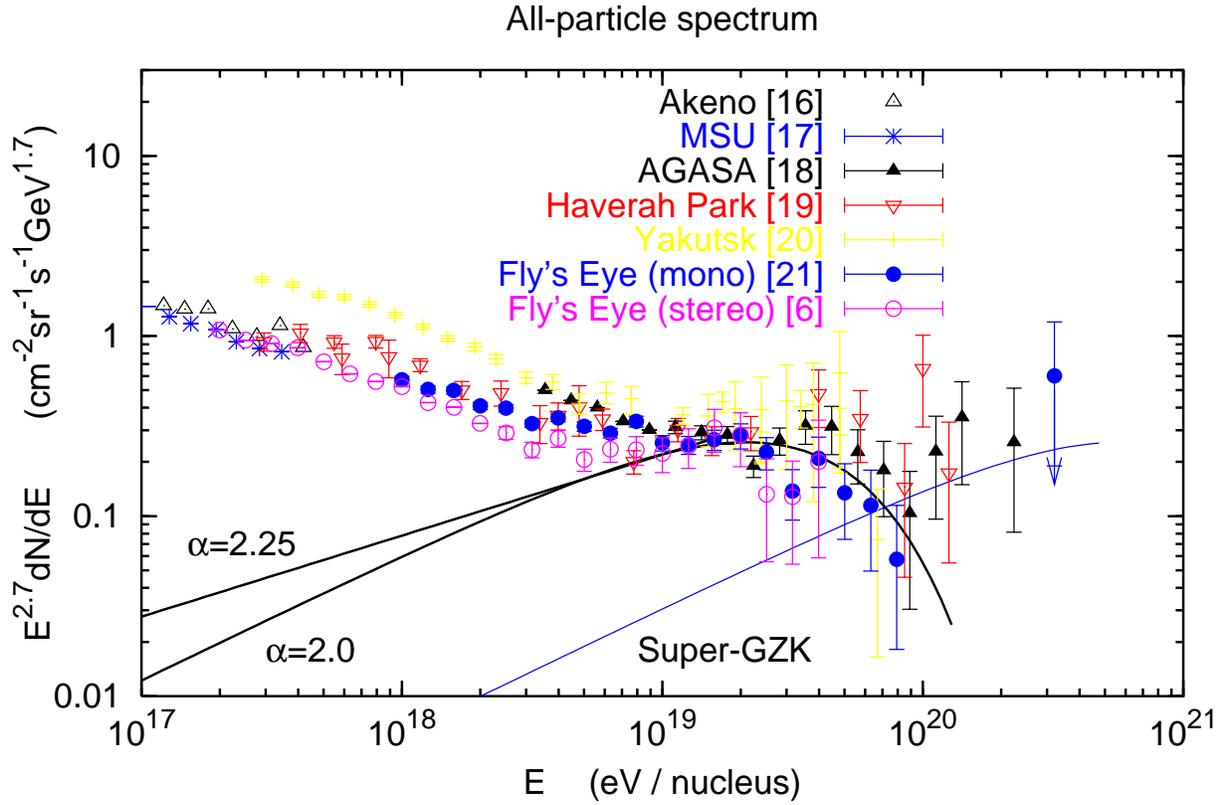}
\caption{Cosmic Ray flux versus cosmic ray energy (extracted from \citep{Gaisser01}).
Fits for two cosmic ray spectral indices ($\alpha$) are shown. The Super-GZK line
refers to a possible contribution from nearby sources assuming $\alpha=2$ (see
\citep{Gaisser01} for details.)}
\label{fig:cr}
\end{figure}

Galactic and extra-galactic high-energy cosmic rays 
are observed at the Earth and in space through
indicators such as synchrotron emission and gamma-radiation.
Some of the most spectacular sites for their origin are the double-lobed radio
sources associated with Active Galactic Nuclei.
Figure \ref{fig:cr} shows a compilation \citep{Gaisser01} of the observed cosmic ray 
spectrum  observed at the Earth.

These high-energy cosmic rays interact with radiation or matter
at the acceleration sites, in transit through intergalactic and
interstellar space, and in the Earth's atmosphere.
Often the result of this interaction is the production of 
pions, kaons, and other particles that decay into muons and an associated
muon neutrino. 
The muon will usually decay into an electron, an electron neutrino and
a muon neutrino.
Each primary cosmic ray interaction typically would produce 
at least three neutrinos and could produce substantially more.

One can estimate the flux of such neutrinos
by models and from observations of the cosmic ray fluxes. 
One has to take into account the cosmic ray interactions 
which at high energies and from distant
sources are primarily with photons.
These estimates and fluxes must satisfy the limits set by the
observations of cosmic rays.  Another possibility to define the
neutrino spectrum is given by the assumption that the high energy gamma-rays
are produced by neutral pion decay. Under this assumption the high energy
gamma-ray spectrum can be used to estimate the neutrino spectrum
from charged pion decay.
In this paper we estimate the muon neutrino and secondary muon flux 
from diffuse and point sources. The sources considered are the ones
which are predicted to produce high energy (around $10^{19}$ eV) 
cosmic rays.
Most models within the standard model of particle physics predict
a muon neutrino rate that is at least twice the electron neutrino
rate and even higher than the tau neutrino rate. The detector acceptance
for muons is also higher than that for electrons and taus since muons can
be detected even when created outside of the detector volume.
For these reasons we
only concentrate on the muon neutrino rate. Here we do not account for the 
flux due to exotic and beyond standard model possibilities 
but leave those topics for a later paper. A review of neutrino physics
within and beyond the standard model can be found in \citep{learned}.

\begin{sloppypar}
There is strong evidence that neutrinos oscillate from one flavor to another
\cite{sk} and that most likely muon neutrinos oscillate into tau neutrinos \cite{sktau}.
Although this effect will lower the muon neutrino rate and enhance the one for tau neutrinos
it is not taken into account in this work. We determine the most optimistic rates which
are for muon neutrinos with no oscillation.
\end{sloppypar}

After reviewing estimates of neutrino fluxes we expand the previous work by including the
fundamental physics aspects of instrument performance. This is done in a generic
way that is independent of the detector configuration. We first determine the 
experimental sensitivity of an ideal detector of km$^2$ incident area to astrophysical
sources and then translate this result to different detector geometries. Our results
are compared to the sensitivity quoted by current and proposed experiments. 

\begin{sloppypar}
In Section~\ref{sec:nuflux} we review and examine estimates of high energy neutrino fluxes  
and upper limits from diffuse and point sources. 
We analyze Sagittarius (Sgr) A East as an example of a
galactic point source and Gamma Ray Bursts and Active
Galactic Nuclei as extra galactic point sources. These are good examples of
the brightest sources.
This section reviews the work of Waxman and Bahcall \citep{wax,WB1,WB2},
of Mannheim, Protheroe and Rachen \citep{mpr}
and the recent estimates by Gaisser \citep{Gaisser01}.
In Section~\ref{sec:rates} we compile and compute the interaction rates, that is, 
the neutrino interactions per unit volume per unit time and 
the number of muons entering or appearing in a generic detector.
This work is directly compared to estimates by Gaisser \citep{Gaisser01}. 
We then expand work previously done and determine the experimental 
sensitivity of an ideal detector to the muon rate.
This is done in a generic form such that in Section~\ref{sec:sens-real} it
is translated to different detector geometries. 
Realistic event rates are then determined for different proposed 
detectors such as IceCube, AMANDA, ANTARES and NESTOR.
\end{sloppypar}

\section{Neutrino Fluxes}
\label{sec:nuflux}
\subsection{Diffuse Flux}
\subsubsection{Waxman and Bachall Limit}
\label{sec:WB}

Waxman and Bahcall (WB) \citep{WB1,WB2} pointed out that the observed 
cosmic ray flux  at high energies implies an upper bound 
on the high-energy astrophysical neutrino flux. 
The latter is produced by the parent cosmic ray particles through pion production. 

This argument holds for sources that are ``optically thin'' 
to the primary cosmic rays. 
``Optically thin" sources are those for which the majority 
of the protons escape and only a fraction interact inside the source. 
Observations of primary cosmic ray flux 
then set a limit on the cosmic production rate of high-energy protons and 
in turn on the production rate of neutrinos.

Waxman and Bachall also account for the 
cosmological evolution of the source activity and redshift 
energy loss of neutrinos due to cosmological expansion.
If the cosmic ray acceleration sites were much more active in the past
(billions of years) than in the present (last 100 million years),
then the flux of ``cosmological'' high-energy neutrinos could be enhanced.

In determining a limit for the diffuse neutrino flux, 
WB assume that all the proton energy is transferred to the pion 
when actually this energy transfer is typically about 20\%. 
Due to this factor, they point out that 
their upper bound exceeds what can be observed by at least a factor of five.

One can conclude that if the WB limit holds, 
the neutrino flux upper bound can guide neutrino telescope designs. 
Neutrino telescopes should be designed to detect realistic fluxes 
which would be at a level well below the WB limit.

The key parameter in setting an upper limit for the neutrino flux from
the collection of ``optically thin'' sources, is the primary cosmic ray
spectral index (which is the power in the power law energy distribution). 
For the relativistic shocks needed to produce 
the very highest energy cosmic rays the spectral index is  
in the range -2 to -2.5 \citep{bell,bland}.
Either will adequately explain what appears to be the extra-galactic 
component of the cosmic rays as shown in Figure~\ref{fig:cr}.

Waxman \citep{wax} has shown that the cosmic ray energy spectrum for energies between $10^{19}$
and $10^{20}$ eV is consistent with what is expected from a homogeneous cosmological
distribution of cosmic ray sources and constrain the spectral index to be in the
1.8 -- 2.8 range. 

Fixing the spectral index to 2, WB \citep{WB1} determine the limit 
on muon neutrino plus muon anti-neutrino extra-galactic flux to be

\begin{equation}
\left( \frac{d\phi_\nu}{dE} \right)_{\rm limit} = \frac{1 {\rm ~to~ } 4 \times 10^{-8} }{E^2} ~~ 
{\rm GeV \,cm^{-2} s^{-1} sr^{-1}} 
\end{equation}
where $\phi_\nu$ is the neutrino flux. Note that this limit is valid for sources
that accelerate cosmic rays to above $10^{19}$ eV.
The range in the coefficient depends upon what evolution is assumed.
For this discussion we use the highest value and obtain an upper limit 
on the muon neutrino flux of

\begin{equation}
\left(\frac{d\phi}{d ln E} \right)_{\rm limit} = \frac{4 \times 10^{-8} }{E} ~~ 
{\rm cm^{-2} s^{-1} sr^{-1}} 
\end{equation}
where E is given in GeV.

\begin{figure}
\plotone{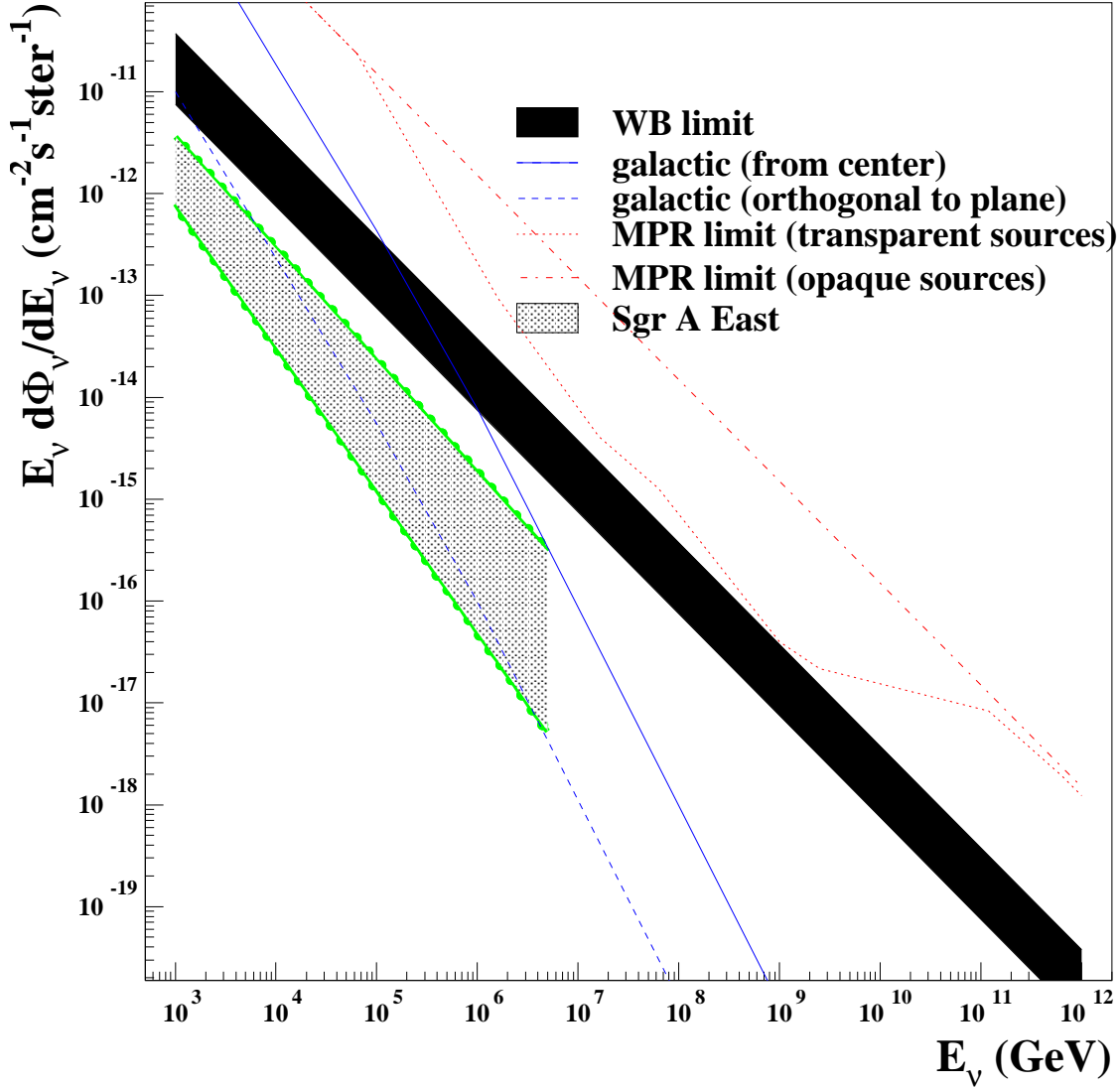}
\caption{Differential neutrino fluxes weighted by neutrino energy from astrophysical 
sources and upper bound on the
total diffuse neutrino flux. The lower edge of the dark shaded area is the WB limit
\citep{WB1} with no cosmological evolution, the upper edge is the same limit 
taking
evolution into account. The dashed line is the MPR \citep{mpr} limit for transparent
sources and the dotted dashed line for opaque sources. The continuous line
is the flux from the Galactic center and the dashed line is the galactic
flux from the direction orthogonal to the galactic plane \citep{IT}.
The light shaded area is the neutrino flux from Sgr A East \citep{crocker} where
the upper edge is the best case scenario flux and the lower edge is the worst
case scenario (see text).} 
\label{fig:astro}
\end{figure}

The WB limit is shown in Figure~\ref{fig:astro}.

\subsubsection{Mannheim, Protheroe and Rachen limit}
\label{sec:mpr}

Mannheim, Protheroe and Rachen (MPR) \citep{mpr} determine an upper limit for
diffuse neutrino sources in almost the same way as WB but with one important
difference: 
they do not assume a specific cosmic ray spectrum but use the
experimental upper limit on the extra-galactic proton contribution. 
While WB base their calculation on a cosmic ray flux 
with a single spectral index equal to -2,
MPR define their spectrum based on current data at each energy.

MPR also extend their calculation for sources that are opaque
(``optically thick'') to nucleons. 
For these sources they set an upper limit using the observed diffuse 
extra-galactic gamma-ray background assuming that the dominant part 
of the emitted gamma-radiation is in the 
Energetic Gamma Ray Experiment Telescope \citep{egret} range.

Their results are also shown in Figure~\ref{fig:astro}. 
Their limit for transparent (``thin'') sources is approximately the same 
as the WB limit for energies 
between $10^7$ and $10^9$ GeV and higher otherwise. 
Their limit allows the rates to be within the area defined by
opaque and transparent sources. 
However, one should bear in mind that fluxes from opaque sources are difficult
to produce. 
The interaction target in these sources must be optimized to allow interactions
with most of the nucleons and at the same time allow pions and muons to decay.
They also require an extraordinary larger energy budget 
than optically thin sources since a higher flux requires more energy. Also
the opacity cuts down the flux of protons before they reach useful high energies implying a much larger initial flux and energy budget 
than  the simple order of magnitude more flux would imply.
As MPR state in their work, the WB limit is closer to current cosmic rays and 
neutrino production models. 
The opaque sources are in the ``hidden'' sources category.

\subsubsection{``Hidden'' Sources}
\label{sec:hidden}

The energy spectrum from opaque sources is not constrained by the
observed cosmic ray flux (see section~\ref{sec:WB}). 
Therefore models which predict such spectra are not limited by the 
WB derivation nor by the MPR for ``thin'' sources.
These models assume sources that are ``optically thick'' to nucleons.
The nucleons must first be accelerated to high energies 
and then encounter a target (radiation or matter) abundant enough
to interact with most of the protons but low enough 
that the pions and muons are able to propagate freely and decay and 
then the neutrinos be allowed to escape.
Berezinsky and Dokuchaev \citep{BD} have proposed such a model
and find that such a source could produce up to 10 muons 
crossing a one square kilometer area per year.
One could argue that such a high flux would be limited 
to a short term (10 years out of billions) outburst.  
However, the long term limit for all sources 
is close to the WB limit. 

Another model that assumes a source which is
``optically thick'' to nucleons has been proposed by Stecker et al 
\citep{stecker}.
It proposes that protons are accelerated to high energies in the AGN core.
These protons produce neutrinos through photo-meson production.
The neutrino flux predicted by this model is shown in 
Figure~\ref{fig:wbagn}. A more recent version of this model \citep{stec96} 
predicts a slightly different neutrino spectrum coming
from quasars. Both predictions expect a  neutrino flux at the level of
the published AMANDA B-10 lower limit \citep{hill}.

A way to avoid both the WB and MPR limits is the
production of neutrinos in a way other than the photo-meson or
proton-nucleon interactions. 
There is a vast list of models that can account for this possibility. 
A list of them can be found in \citep{WB2}. 
Most of these involve new or ``exotic'' physics and we do not
include them as ``astrophysical'' sources.

As discussed in section~\ref{sec:mpr}, neutrino fluxes from hidden sources
are less likely to be produced.

\subsubsection{GZK Fluxes}
\begin{sloppypar}
Very high-energy protons traveling through intergalactic space 
interact with the cosmic background photons and photo-produce pions.
Greisen and Zatzepin and Kuz'min \citep{GZK} first pointed out
that this process would occur and that it would set an upper bound
to the maximum energy (a few times $10^{19}$~eV) for a proton 
traveling intergalactic distances (hundreds Mpc).
\end{sloppypar}

\begin{figure}
\plotone{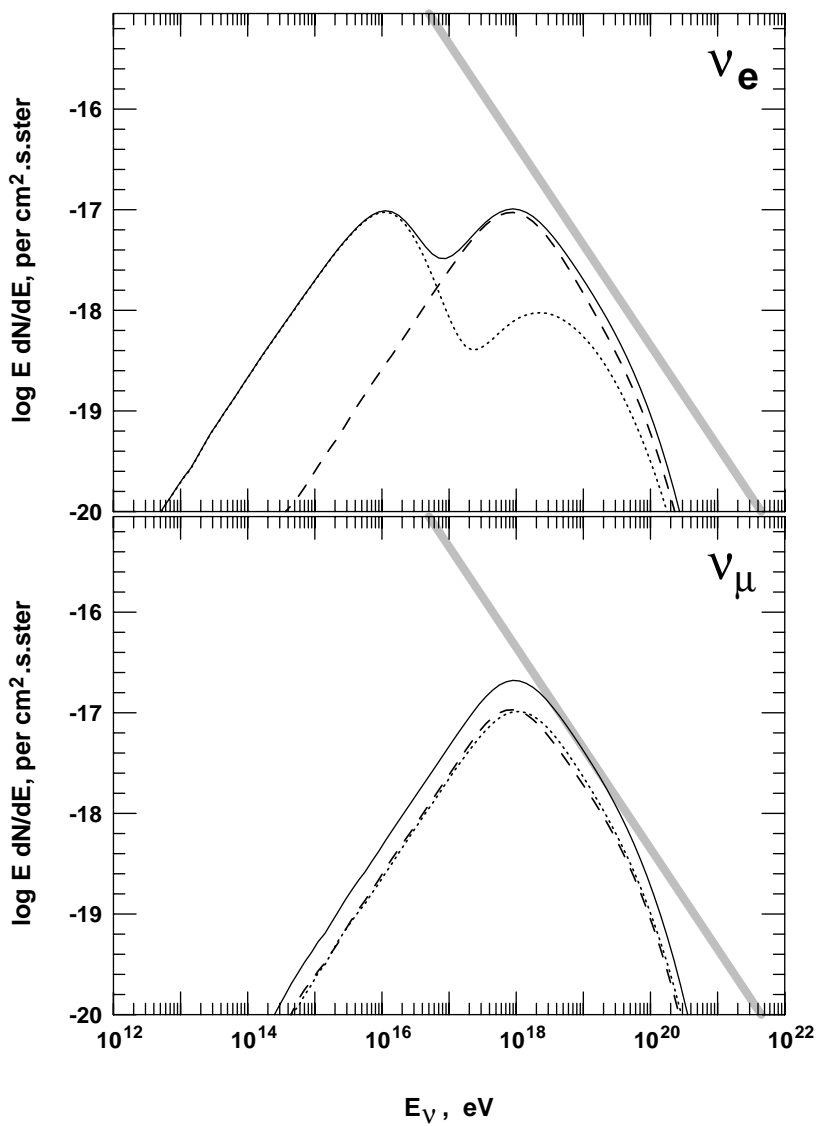}
\caption{Differential fluxes weighted by neutrino energy 
of neutrinos generated by ultra high energy
protons. Fluxes of electron neutrinos (dashed lines) and anti-neutrinos 
(dotted lines) are shown in the upper panel. 
The lower panel shows the fluxes of muon neutrinos and anti-neutrinos. 
Solid lines show  the sum of neutrinos and anti-neutrinos. These
fluxes were calculated in \citep{engel} using 
the same injection power and cosmological source evolution as
the WB limit~\protect\citep{WB1}. This limit is represented by the
shaded line. Figure extracted from \citep{engel}.}
\label{fig:GZKflux}
\end{figure}

The production and flux of high energy neutrinos originated from 
intergalactic propagation of ultra high
energy cosmic rays has been determined in \citep{engel}. 
They assume that the
ultra-high energy cosmic rays are of astrophysical origin and 
show their
result for different assumptions on cosmic ray source distributions, 
injection spectra and cosmological evolution. 
Their results with the same assumption for the injection power and
cosmological evolution as Waxman and Bahcall \citep{WB1}
are shown in Figure~\ref{fig:GZKflux}.

\subsubsection{Neutrinos from the Galaxy}
There is a significant diffuse flux of neutrinos
created by interactions of the galactic and extra-galactic cosmic rays
with interstellar matter and starlight.
The spectrum of cosmic rays is reasonably well known,
as is the distribution of targets in our Galaxy. We call the flux
of neutrinos created by interactions in the interstellar medium 
the ``galactic flux''.

\begin{sloppypar}
Estimates of the galactic neutrino flux have been made by 
Domokos \citep{Domokos}, Berezinsky et al. \citep{Bere}, and 
Ingleman and Thunman \citep{IT}.
The diffuse galactic neutrino flux can be separated from the local 
atmospheric neutrino flux at energies above $10^{15}$ eV.
\end{sloppypar}

Above these energies 
the flux of diffuse galactic neutrinos stays harder
than the atmospheric flux both because the original galactic cosmic rays
have a harder spectrum nearer the center of the galaxy 
and most important the interstellar material is sufficiently thin
and far away that the muons have adequate range and time to decay.

The muon neutrino flux (per GeV cm$^2$ sr s) is estimated \citep{IT} by

\begin{equation}
\phi_\nu = \lbrace \matrix{ 3.0 \times 10^{-6} R E_\nu^{-2.63} ~~~~~ 
E_\nu < 4.7 \times 10^5 ~ {\rm GeV} \cr
1.9 \times 10^{-4} R E_\nu^{-2.95} ~~~~~ E_\nu > 4.7 \times 10^5 ~ {\rm GeV} }
\end{equation}
where $R$ is the distance to the edge of the galaxy in Kpc  and E$_\nu$ is the neutrino
energy in GeV.

The resulting flux from the direction of the  
center of the galaxy (20.5 kpc) and from the
direction orthogonal to the galactic plane (0.26 kpc) are shown in 
Figure~\ref{fig:astro}. Both the WB and MPR limits are for extra-galactic
neutrinos.

One important point when considering the galactic flux is the direction
from which the neutrinos come. 
Experiments in the South Pole will have an additional background
from atmospheric muons when detecting neutrinos from the center of the 
galaxy since all events come from above the horizon. 
The atmospheric muon spectrum is steeper than the atmospheric
neutrino spectrum ~\citep{gaisser}.  The flux of atmospheric muons 
surviving to a depth of 1 km is 200 times larger than the 
atmospheric neutrino induced muons at 1 TeV.  
Furthermore, atmospheric muons that survive to kilometer depths
are often produced in collimated bundles near the core of the 
parent cosmic-ray shower.  Bundles of ~10 TeV muons can easily be
misidentified as a single high energy muon of 100 TeV or more.
The spectral characteristics of muon bundles are not yet well characterized
and lead to poor estimates of sensitivity above the horizon. 
For this reason we assume that atmospheric muons compose a irreducible
background.
In section~\ref{sec:sens} we determine the sensitivity for an expanded version
of ANTARES (expanded to a km$^2$ incident area) to the neutrino flux from the
galactic center.

\subsection{Point Sources}
\label{sec:point}
In this section we consider sources that produce high-energy particles.
Other than the hidden sources mentioned in section~\ref{sec:hidden} active
galactic nuclei (AGNs) and gamma-ray bursters are examples of extra galactic point
sources. Other extragalactic point sources such as blazars \citep{dermer}
(which compose a subclass of active galactic nuclei)
might produce a lower flux of neutrinos when compared to GRBs.
As an example of galactic point sources we show the neutrino
flux from Sagittarius (Sgr) A East.

In principle a sufficiently bright point source
can have their locations determined by the arrival direction of these
particles (or by the particles produced by them) at the detector.
The relevant issue is that, if the location of the source is pre-known,
then the effective atmospheric background is effectively reduced
by the ratio of the effective solid angles. 
The intrinsic angular deviation between the neutrino and daughter muon is 
of order $0.7^\circ/\sqrt{E_\nu/TeV}$ \citep{gaisser} so that the effective 
solid angle is of order one square degree.
If the detector angular resolution is of this order, 
then the relative enhancement of signal-to-background can be as much as $10^{4}$.

\subsubsection{Active Galactic Nuclei}

Active Galactic Nuclei (AGN) are one of the brightest known astrophysical sources.
They produce a multi-wavelength spectrum that goes from radio to TeV gamma rays.
How such high-energy gammas are produced is still to be fully understood.

\begin{sloppypar}
In the more conventional model high energy photons are produced by 
inverse Compton scattering of accelerated electrons on thermal UV photons \citep{compt}. 
This description is supported by multi-wavelength observations 
of Mkn~421 \citep{mkn421} although there are adjustments to be made \citep{halzas}.
\end{sloppypar}

Other models \citep{managn,halzas,protagn} 
describe the production of high energy gamma rays through the
decay of photo produced neutral pions.
These protons would be produced and accelerated in the AGN jet. 
This mechanism would be responsible for the production of the 
observed gamma-ray background.

\begin{figure}
\plotone{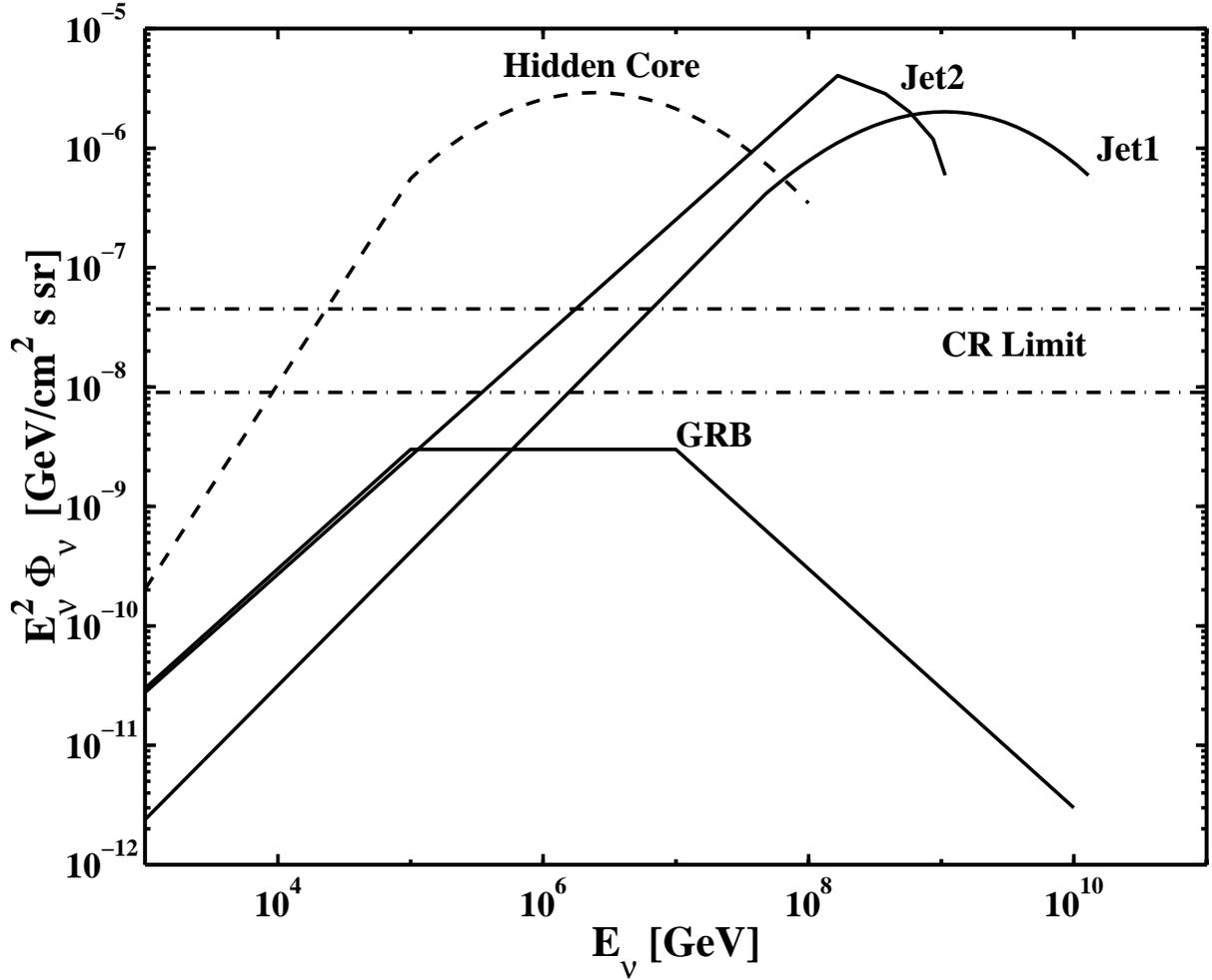}
\caption{Differential muon neutrino flux weighted by the square of the 
muon neutrino energy versus muon
neutrino energy. The WB limit is labeled as CR limit and compared to 
theoretical models. The Jet1 models
refers to results from \citep{managn} and Jet2 from \citep{halzas}.
These models should be constrained by the WB limit. As they violate
the limit, WB conclude that one of their assumptions must be wrong.
The GRB model is a prediction for neutrinos from Gamma Ray Bursts 
\citep{WB1,wbgrb}. The WB limit is also
compared to a hidden source model \citep{stecker} for which the WB limit
does not apply (see section~\protect\ref{sec:hidden}). This model  
is at the level of the published AMANDA B-10 lower limit
(Hill et al. 2001). Figure extracted from
\protect\citep{WB1}.}
\label{fig:wbagn}
\end{figure}

Figure~\ref{fig:wbagn} shows that the non conventional model
\citep{managn,halzas,protagn}
predictions are much higher than the WB limit. 
Waxman and Bahcall \citep{WB2} show that the sources in these models 
are optically thin and therefore should be constrained by their limit. 
They conclude that at least one of the basic assumptions of these models 
is not valid.

It is important to note that the AGN sources are intermittent and that
they might violate the WB limit in a temporary basis. 
However they should not do so when averaged over time.

\subsubsection{Gamma-ray bursts}
\label{sec:grb}

Energetic gamma-ray bursts (GRB) seem to be successfully explained by fireball
models (\citep{piran} and references therein). 
Recent observations suggest that they originate in cosmological sources. 
Waxman and Bahcall \citep{wbgrb} propose that
neutrinos are a consequence of these fireballs. 
They will be produced by  photo-meson production between the fireball 
gamma-rays and accelerated protons.
\citep{wbgrb,WB1} derive the energy spectrum and 
flux of high energy neutrinos created in this way. 
Their result is in agreement with \citep{rachen}.

\begin{sloppypar}
Figure~\ref{fig:wbagn} shows the predicted spectrum of high energy 
neutrinos from GRBs. 
This spectrum is consistent with the WB limit and the intensity over 
$2\pi$ sr coverage would be about 10 neutrino induced muons per year 
in a detector with $km^2$ incident area \citep{WB1}.
These neutrinos have the advantage of spatial and temporal coincidence with
GRB photons 
which can be used as a tool to reduce the atmospheric neutrino background.
The neutrinos produced from hadronic interaction may arrive on a time scale 
of about an hour 
after the photons in the model of simple acceleration \citep{WL}.
\end{sloppypar}

\subsubsection{Neutrinos from Galactic Point Sources}
\label{sec:galpoint}
The neutrino flux from Sgr A East is analyzed as an example of a galactic point
source. This is one of the dominant radio emitting 
structures at the Galactic 
Center and is a supernova remnant-like shell \citep{melia}. It has been shown \citep{melia}
that the highest energy component in the Sgr A East spectrum is compatible with a gamma-ray
spectrum produced by pion decay. This energy spectrum fits the one from the galactic center source
2EG J1746-2852 measured by EGRET. The neutrino flux
from Sgr A East was determined under the assumption that pion decay 
constitutes the gamma-ray source 2EG~J1746-2852 \citep{crocker}.
Pions are produced in proton-proton collisions in
shock regions and the maximum energy of the shocked protons is $\sim 5 \times 10^{6}$~GeV.
Neutrinos are produced from pion and muon decay. There is a ratio of about
67\% muon like to 33\% electron like neutrinos. We determine the sensitivity of a 
km$^2$ incident area to the largest neutrino flux, ie, the muon neutrino flux.

\begin{sloppypar}
If neutrino oscillations occur as expected from  Superkamiokande results \citep{sk} one
should take this effect into consideration \citep{crocker,crocker2}. In this
case, a portion of the muon neutrinos will most likely \citep{sktau} become tau
neutrinos. In this work we do not include oscillations and it represents an upper
limit for muon neutrinos.

In Figure~\ref{fig:astro} we show the muon neutrino flux from Sgr A East as
determined in \citep{crocker}. The band represents
the range between the best case scenario for which the proton energy spectrum will follow a power law
with spectral index equal to -2.1 and the worst case scenario with spectral index equal to
-2.4 \cite{crocker}. The cutoff is due to the maximum energy achieved by the proton.
In section~\ref{sec:sens} we will determine the detectability of this flux of
neutrinos.
\end{sloppypar}

Another galactic neutrino source might be supernova remnants (SNR). The 
CANGAROO collaboration has measured the gamma ray spectrum of a few of
these sources \citep{can1,can2}. If the
gammas were produced as a product of pion decays, neutrinos will also be produced.
The neutrinos from SNR will however have a lower flux than the ones from Sgr A East and
the cutoff energy will be at the TeV level instead of at the PeV level \cite{crocker2}. 

The brightest cosmic X-rays sources -- the X-ray binaries -- might also produce high
energy neutrinos. Composed by compact objects such as black holes or neutron stars,
they accrete matter from their companion stars. The accretion process might accelerate
protons to high energies. The interaction of these protons with either the accreted matter 
or matter from companion stars will produce a neutrino flux. The neutrino flux from
microquasars, which are Galactic jet sources associated with some classes of X-ray binaries
has been determined in \citep{waxqua}.

\subsubsection{Neutrinos from the Sun}

The high energy neutrino flux originating from cosmic ray interactions with 
matter in the Sun has been determined in \citep{itsun}. Although it is
higher than the atmospheric neutrino flux, the conclusion is that the 
absolute rate is low. Within the Sun's solid angle, the neutrino energy spectrum
will follow the atmospheric neutrino spectrum but is about 3 times higher.  
According to these authors the low rate precludes the
Sun as a ``standard'' candle for neutrino telescopes and also limits
neutrino oscillation searches. A detailed analysis of the influence of
neutrino oscillations on the high energy neutrino event rates from the Sun 
can be found in \citep{sun}.

\subsection{Atmospheric Neutrino Fluxes}
\label{sec:atm}

The atmospheric neutrino flux is the main background for neutrinos from
astrophysical sources.
As a parameterization of this background flux we use the derivation made by
Volkova \citep{volkova} as described also in \citep{alb}.

Volkova derives the atmospheric neutrino flux from the decay of light mesons
($K, \pi$) and muons and from the decay of short-lived particles 
(prompt decay) which mainly includes charm particles. 
The latter will only be significant at higher energies (above a PeV).

\begin{sloppypar}
We compare this flux with that obtained by \citep{honda} and 
\citep{agrawal}.
The biggest difference between these calculations is of about 15\% 
(see Figure~14 of ref.~\citep{honda} and Figure~7 of ref.~\citep{agrawal}). 
In the energy range of interest to our work, the discrepancy is less
than 15\% and the Volkova
spectrum is underestimated in relation to these other spectra. 
\end{sloppypar}

Since the atmospheric neutrino flux is irreducible, understanding
the effect of its uncertainty is important. The uncertainties come from
the primary cosmic ray flux measurement and from 
the inclusive cross section for proton -- nucleon interactions.
We assume the biggest discrepancy between the atmospheric flux calculations
(15\%) as the uncertainty in the magnitude of the atmospheric flux.

Since the uncertainty in the primary spectrum increases with energy
there is also an uncertainty in the slope of the spectrum. At
lower energy there are more data and the uncertainty 
is mainly due to instrumental efficiency and exposure factor. 
At higher energies the uncertainty is dominated by limited statistics.
In general \citep{honda,agrawal} the uncertainty in 
the spectrum slope is assumed to be about 10\% below
3 GeV increasing to 20\% at $3 \times 10^3$ GeV and remaining constant from
thereon. 
As discussed also in \citep{honda,agrawal} the uncertainties in the interaction
model at higher energies is estimated around 10\%. 
Neutrino oscillations do not affect the atmospheric spectrum at the energies
of interest here.

One can therefore estimate the overall uncertainty 
in the atmospheric neutrino flux as around 20\%. 
We will show that this uncertainty does not affect 
significantly our results. 

Although the atmospheric neutrino flux is always considered as a background for
neutrino astrophysics, the fact that the flux at higher energies is not well
determined by experiments, makes it an important measurement to be performed 
with neutrino telescopes. 
It has been suggested that the standard model neutrino
cross section above $10^8$ GeV might be lower than expected \citep{dicus}.
Measuring the atmospheric neutrino flux at these energies can determine
the neutrino -- nucleon cross section \citep{weiler}. It is important to note
that these energies cannot be directly probed by accelerator experiments.

\subsection{Summary of Fluxes}
\label{sec:summary}

\begin{figure}
\plotone{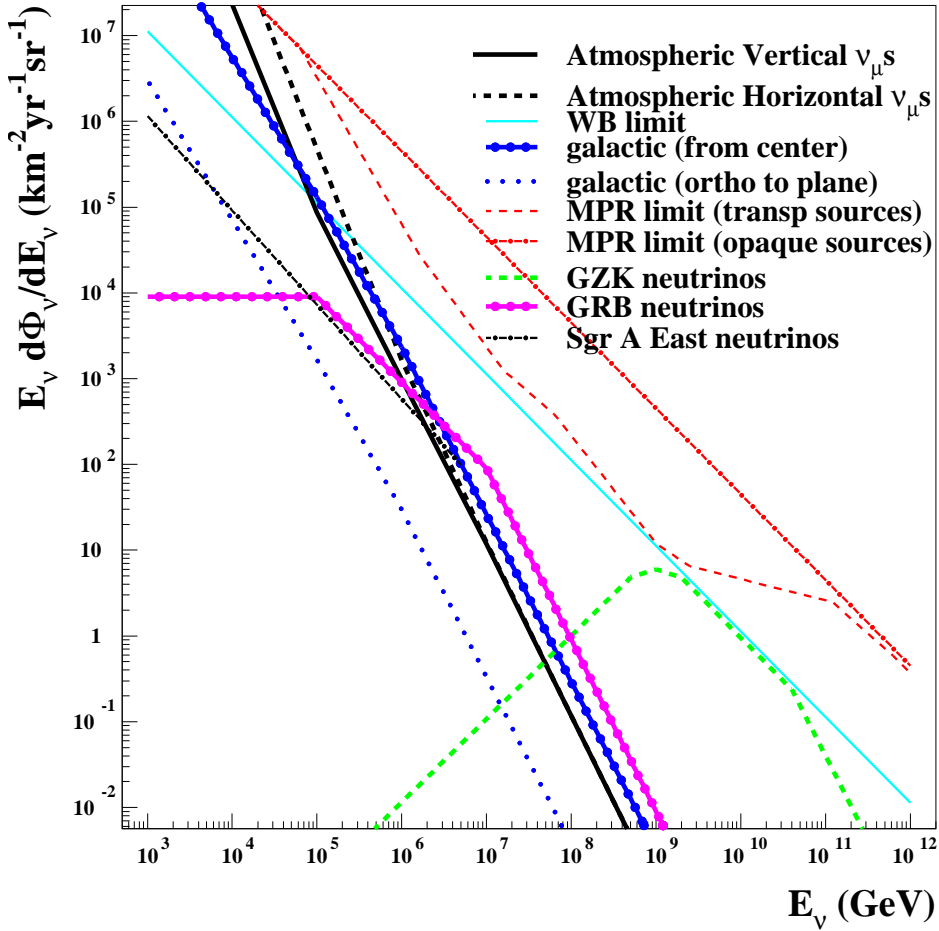}
\caption{Differential muon neutrino fluxes weighted by neutrino energy 
summary containing (1) Atmospheric vertical
(solid line) and horizontal (dashed) 
muon neutrino energy spectra based on \protect\citep{volkova}. 
The reason why these two fluxes overlap at high energies (above 
$10^6$ GeV) is that prompt decays (mainly from charm particles) 
dominate
the atmospheric flux when compared to meson and muon decay at these
energies.(2) WB \protect\citep{WB1} limit (with evolution) 
for neutrino fluxes from astrophysical sources (large solid line).
(3) Galactic neutrino flux \protect\citep{IT} from galactic center 
(dotted dashed line) and from direction orthogonal to the galactic 
plane (dotted).
(4) MPR \protect\citep{mpr} limit
for neutrino fluxes from sources
transparent to nucleons (medium dashed line) and from sources 
opaque to nucleons (medium dotted dashed).
(5) Estimate of GZK neutrinos \protect\citep{engel} (large dashed line).
(6) Estimate of GRB neutrinos \protect\citep{wbgrb} (large dot dashed line)
(7) Estimate of Sgr A East neutrinos \protect\cite{crocker} assuming the
best case scenario (see text) (medium dot dashed line).}
\label{fig:sum}
\end{figure}

Figure~\ref{fig:sum} shows a summary of the expected fluxes arriving
at the surface of the Earth from sources which produce high energy (around
$10^{19}$ eV) cosmic rays.  The atmospheric background dominates
at low energy. Neutrinos from the center of the galaxy  
contribute a small excess to the atmospheric flux at energies above 
$10^6$ GeV, but are constrained by a smaller solid angle.
They are not in the field of view of South Pole detectors, but are
interesting for more equatorial detectors.  
Diffuse limits on astrophysical neutrinos are characterized by the WB limit.  
The WB limit is avoided by the MPR limits as they abandon the power-spectrum
for optically thin limit and propose ``hidden'' source contributions
for the optically thick limit.
GZK fluxes are shown at the highest energies.
The most promising flux to be measured is the one from GRBs since
it allows reduction of the background 
through knowledge of arrival time and direction as well as the one from
Sgr A East.

Detected event rates are smaller since the 
neutrino conversion into a muon and the efficiency of the detector
have to be taken into account.  Event rates and experimental sensitivity
are considered in the next section.

A sensible design criteria for a neutrino detector is that
it be sensitive to the
highest known neutrino flux from astrophysical sources, namely five
times below the WB limit.  With such a sensitivity, backgrounds
can be characterized and new diffuse fluxes discovered.
Physics measurements of the new fluxes, namely brightness, energy spectrum
and points on the sky, will require substantially more events. 

\section{Event Rates and Sensitivity}
\label{sec:rates}
Neutrinos cannot be directly detected, because they do not 
deposit a significant amount of
energy as they pass through matter.  For a neutrino to be observed
it must undergo an electro-weak interaction with another particle 
resulting in detectable secondaries.  
 Since the neutrino interaction cross section is small, the probability
of an interaction can be increased if 
large, dense targets are used, such as the Earth.

The neutrino nucleon cross section increases as a function of energy 
resulting in two effects.  
First, high energy neutrinos are more likely 
to interact within the detector volume.  
Second, the flux of neutrinos that reach the detector is attenuated at high 
energy because of neutrino interactions in the Earth.
High energy muons travel many kilometers before stopping or decaying.
The advantage of detecting muons is that the detector could be sensitive to 
neutrino interactions over a length equal to the muon range.
Unfortunately, because high-energy muons lose energy rapidly and 
because at low
energy there is a large atmospherically-produced neutrino background,
this potential gain is reduced. 

Convolution of the neutrino flux, conversion cross sections, muon range
and deposited energy is the subject of the next sections.

\subsection{Neutrino Interactions}
The flux of leptons converted from the incident neutrino flux is
\begin{equation}
\phi_{lepton} = \phi_\nu P(\nu \rightarrow lepton)
\end{equation}
where $ P(\nu \rightarrow lepton)$ is the probability that a neutrino
suffers an interaction and produces a lepton.  
This probability is given by
\begin{equation}
P_I = P(\nu \rightarrow lepton) 
    = \int_0^{path} {\rm n} \sigma(\nu \rightarrow lepton) dx
\label{eq:pi}
\end{equation}
where ${\rm n}$ is the nucleon number density and $\sigma$ is the neutrino nucleon
cross section and the path is the distance the neutrino traveled.
In the following, we abbreviate the notation:
$P_{\rm I} = n \sigma_{\nu} L$ where $L$ is the neutrino path.

The deep inelastic neutrino cross sections, $\sigma_\nu$ 
are determined using CTEQ4-DIS parton distribution functions as described 
in \citep{gandhi98}.
In the case of muon neutrinos, the interaction
probability is dominated by the charge current (CC) cross section, 
but there is also some degradation of
neutrino energy by the neutral current (NC) cross section.

Neutrino fluxes will suffer attenuation as they pass through the Earth.
The differential flux is given by
\begin{equation}
\frac{d\phi_\nu}{dx} = -n \sigma_{\nu} \phi_\nu
\label{eq:dfatt}
\end{equation}
where 
$x$ is the distance traveled by the neutrino. 
Integrating over the path length traversed by the neutrino, we find 
\begin{equation}
\phi_\nu = \phi_{\nu 0} e^{-\int n \sigma(\nu) dx }  
         = \phi_{\nu 0} P_S
\label{eq:attfl}
\end{equation}
where $\phi_{\nu 0}$ is the flux at the earth's surface, $P_S$ is the
survival probability and the argument of the exponential term is $P_I$
integrated over all cross sections that make the neutrino cease to exist.

\begin{figure}
\plotone{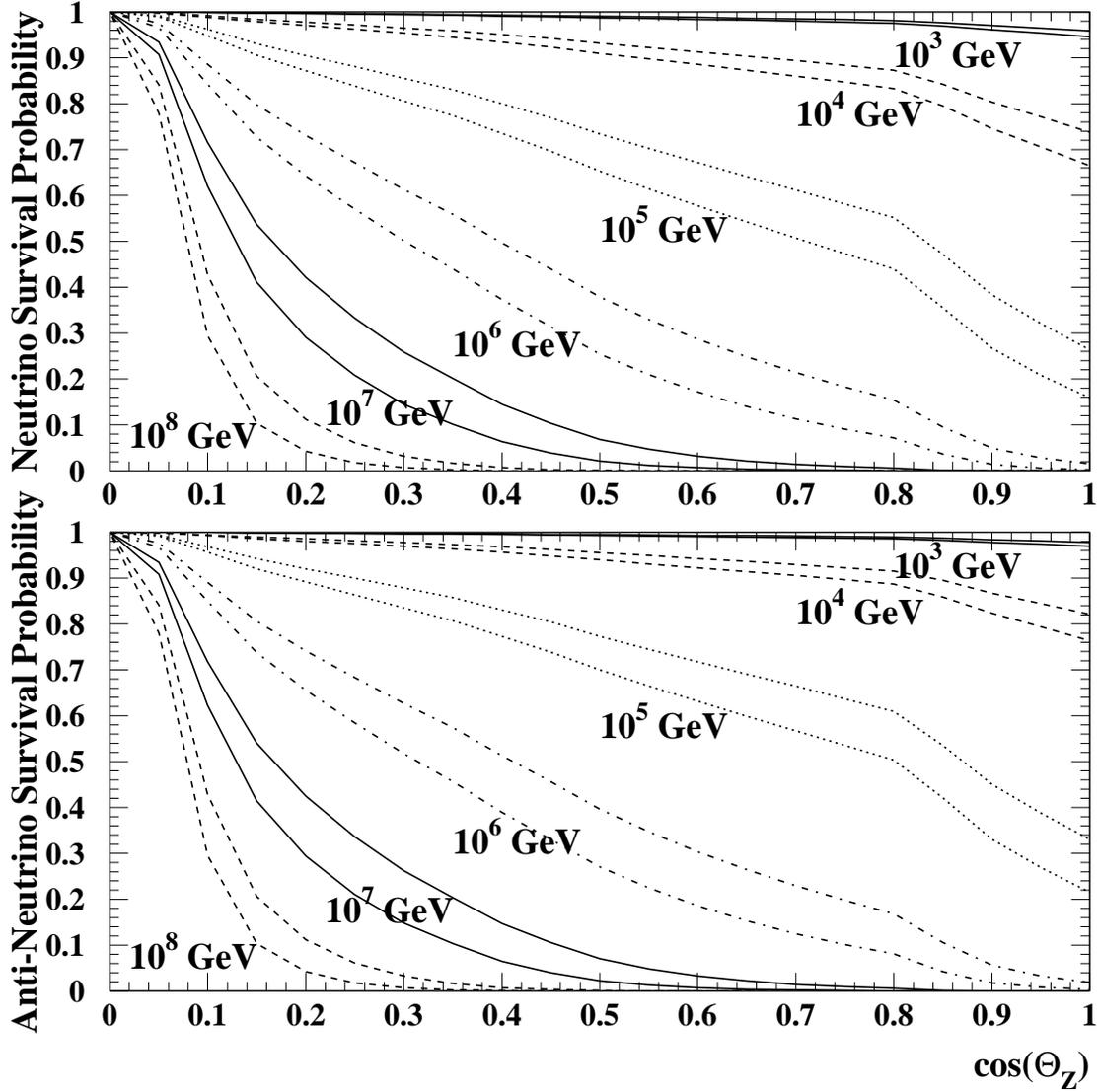}
\caption{Survival probability for neutrinos and anti-neutrinos
transversing the Earth as a function of $\cos(\theta_Z)$.
The horizon is at 0, and  a path through the center of the earth is at 1.  
The upper lines include only CC interactions, the lower lines include both
CC and NC interactions.
Differences due to NC interactions have a maximum effect at high energy
where the distributions drop sharply.}
\label{fig:survival}
\end{figure}

Figure \ref{fig:survival}
shows the muon neutrino survival probability at a point on the 
surface of the earth for a variety of neutrino energies as a
function of the cosine of the earth angle, $\theta_Z$
\footnote{$\theta_Z$ is the angle from local zenith. An upward going neutrino,
that is, one coming from the direction of the center of the Earth as viewed 
from the detector has a zenith angle of zero degrees.}.
The number density (see Equation~\ref{eq:pi}) 
is determined by an integration
of the earth density profile which is taken from \citep{gandhi96,earden}.  
The upper lines are the probability using
only CC interactions.  The lower lines include both CC and NC interactions.
It shows that the NC interaction will decrease the lepton flux by about 10\%.
At high energy, the steep cut-off results in as much as a 20\% decrease.
The NC interaction does not actually remove the neutrino from the flux,
but degrades its energy.  We include it in our analysis to be conservative,
and treat the CC-only case as an upper bound to the systematic 
uncertainty.

From Figure~\ref{fig:survival} one can see that 
the survival probability becomes quite small for high energy neutrinos
due to significant attenuation.  At low energies, the attenuation
is negligible, but it begins to contribute above about 10 TeV.
The Earth becomes essentially opaque to neutrinos at energies of above a PeV
\citep{gandhi98}.

The effect of the dense 
Earth core can be seen near $\cos{\theta_Z}$ of 0.8. 
Neutrinos that travel close to the Earth axis will go 
through most of the Earth core
and therefore increase their probability of interaction.
There is also evidence
of a thin crust at small $\cos{\theta_Z}$ where neutrinos
only pass through the crust.

The flux of muons interacting in a detector volume is given by
\begin{equation}
\phi_{lepton} = \phi_{\nu 0} \times P_S \times P_I 
\label{eq:ncont}
\end{equation}
where the incident flux, $\phi_{\nu 0}$, is the flux in Figure~\ref{fig:sum}
and $P_I$ is now integrated over the neutrino path through the detector.

\subsection{Muon Fluxes}
Muons will propagate, losing energy until they eventually come to 
rest and decay or undergo a nuclear reaction.  
The mean energy loss is given by 
\begin{equation}
\frac{dE_\mu}{dx} = a + b E_\mu
\end{equation}
where asymptotically $a = 2 \times 10^{-3} {\rm GeV cm^2 / g}$ and 
$b = 4 \times 10^{-6} {\rm cm^2 / g}$ \citep{pdg} are
respectively ionization and radiation loss parameters.

Approximating the parameters $a(E_\mu)$ and $b(E_\mu)$ as constants, 
simple integration over the muon path yields the muon range, $R_\mu$.
\begin{equation}
R_\mu = \frac{1}{b} \ln \left( 1 + \frac{b}{a} E_0 \right)
\label{eq:muran}
\end{equation}
where $E_0$ is the muon initial energy. 
Below the critical energy \footnote{Critical energy is the one for which the
probability for a nuclear interaction in one nuclear mean free path equals the 
decay probability in the same path. It is given by $E_{critical} \equiv \frac{b}{a}$.}
ionization losses dominate and the muon range is
well described by the average energy loss.  
At high energies where radiative processes dominate, large fluctuations
develop and a stochastic approach is needed.
The mean muon range is shorter than Equation~\ref{eq:muran}
would suggest.
Lipari and Stanev \citep{lipari} developed a Monte Carlo
with the purpose of taking these fluctuations into account.  
We use their Monte Carlo to determine the muon range.
This approach also accounts for the energy dependence
of the energy loss parameters $a$ and $b$.

\begin{figure}
\plotone{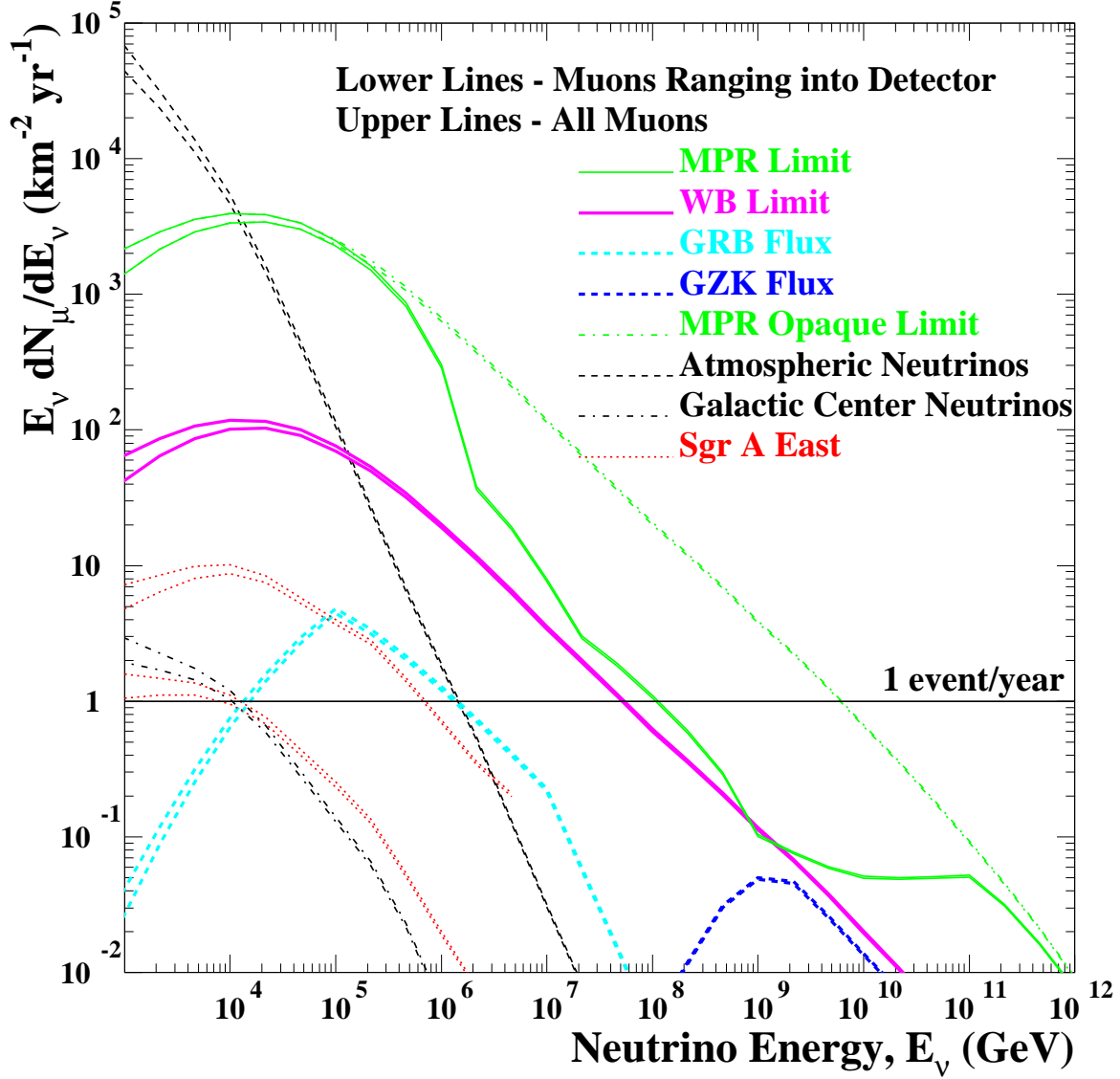}
\caption{Differential upgoing muon and anti-muon rate weighted by neutrino energy 
as a function of neutrino energy.
Most muons range into the detector although the fraction starting
inside increases at low energy. The horizontal line shows where 
1 event/year is expected for each half decade in energy.}
\label{fig:muflux}
\end{figure}

We now determine the upgoing muon rate in an idealized detector located
1.5 km below the surface of the earth with a detector path length of 1 km
and a km$^2$ incident area for all incident angles. Two muon rates are relevant, those 
where the muon is produced inside the detector 
($P_I$ integrated over 1 km) and those where the muon originates outside the 
detector and ranges into the detector ($P_I$ integrated over the muon range).  
Figure~\ref{fig:muflux} shows the rate of upgoing muons plus anti-muons 
weighted by neutrino energy 
versus neutrino energy for the WB limit, the MPR limit, GRB flux, GZK flux,
and for atmospheric neutrinos.  
Only muons with more than 100 GeV are included.  
These results are in good agreement with \citep{Gaisser01}.
The lower curve is the flux of muons plus anti-muons that start 
outside the detector volume and range into it.  
The upper curve includes the flux of
muons and anti-muons that start in the detector volume.
Except for very low energies,
the dominant flux is from muons that range into the detector.

Muons associated with the galaxy are not isotropic. 
We make estimates for a detector at a Mediterranean latitude of +35 degrees
corresponding to a expanded version of the ANTARES or NESTOR detector.  
A North Pole detector would see a rate 1.6 times higher. The galactic center
is above the horizon at the South Pole where the atmospheric muon background
is too high. Figure~\ref{fig:muflux} shows the
rate of neutrinos from 75 square degrees in the galactic plane, centered
on the galactic center.  The rate is much smaller than the Atmospheric 
neutrinos in the same solid angle.  We conclude, therefore that 
Galactic neutrinos will not be detectable.

We also determine the muon rate for Sgr A East at a latitude of 35 degrees. 
Two rates are shown in Figure~\ref{fig:muflux} corresponding to hard and 
soft proton spectra with a cut-off at $5\times10^6$ GeV (see 
section~\ref{sec:galpoint}).  Background neutrinos from the galaxy 
in 1 square degree surrounding Sgr A East are about 75 times smaller than
the rate shown in the figure.
For point sources such as GRB and Sgr A East, 
the primary background of atmospheric neutrinos is reduced in 1 square 
degree by a factor of about $5\times10^{-5}$ 
again leaving a nearly background-free source detection.

Muon rates are substantially lower than the neutrino rates shown earlier.  
The line on Figure~\ref{fig:muflux} shows where
1 event/year is expected for each half decade in energy.  
The GZK flux peaks at $10^9$ GeV, just below the WB limit.  
About 0.17 muon per year is expected from the GZK flux.

\begin{sloppypar}
For all but the atmospheric flux, equal numbers of neutrinos and
anti-neutrinos are expected.  Atmospheric neutrinos
exceed anti-neutrinos by a factor of 2.5 at 1 TeV \citep{agrawal}.
By about 100 TeV, however, the neutrino and anti-neutrino cross 
sections become equal.  Across the whole spectrum, the uncertainty due 
to muon/anti-muon composition can be neglected since it is smaller than 
the theoretical uncertainty of $\pm20$\%.
\end{sloppypar}

Since the neutrino energy is not measured by the detector, 
but rather the muon energy is estimated by energy deposition in the detector,
the results are better shown as a function of the muon energy estimation.
The latter can be achieved in three steps: (1) first the flux as a function of
the muon energy,
(2) as a function of the muon energy at the detector, and 
(3) finally flux as a function of the energy deposition in the detector.
The muon flux as a function of muon energy is given by
\begin{equation}
\phi_\mu = 
\int_0^\infty \frac{d}{dE_\mu} (\frac{d\phi_\nu}{dE_\nu}) dE_\nu  
\end{equation}

\begin{sloppypar}
The average energy loss in a CC interaction, $y = (1-E_\mu/E_\nu)$, 
for neutrinos energies 
between 10 GeV and 100 GeV is 0.48 gradually
decreasing to about 0.2 at high energies \citep{gandhi96}. 
At high energy, the muon gets about 80\% of the neutrino energy.  
\end{sloppypar}

\begin{figure}
\plotone{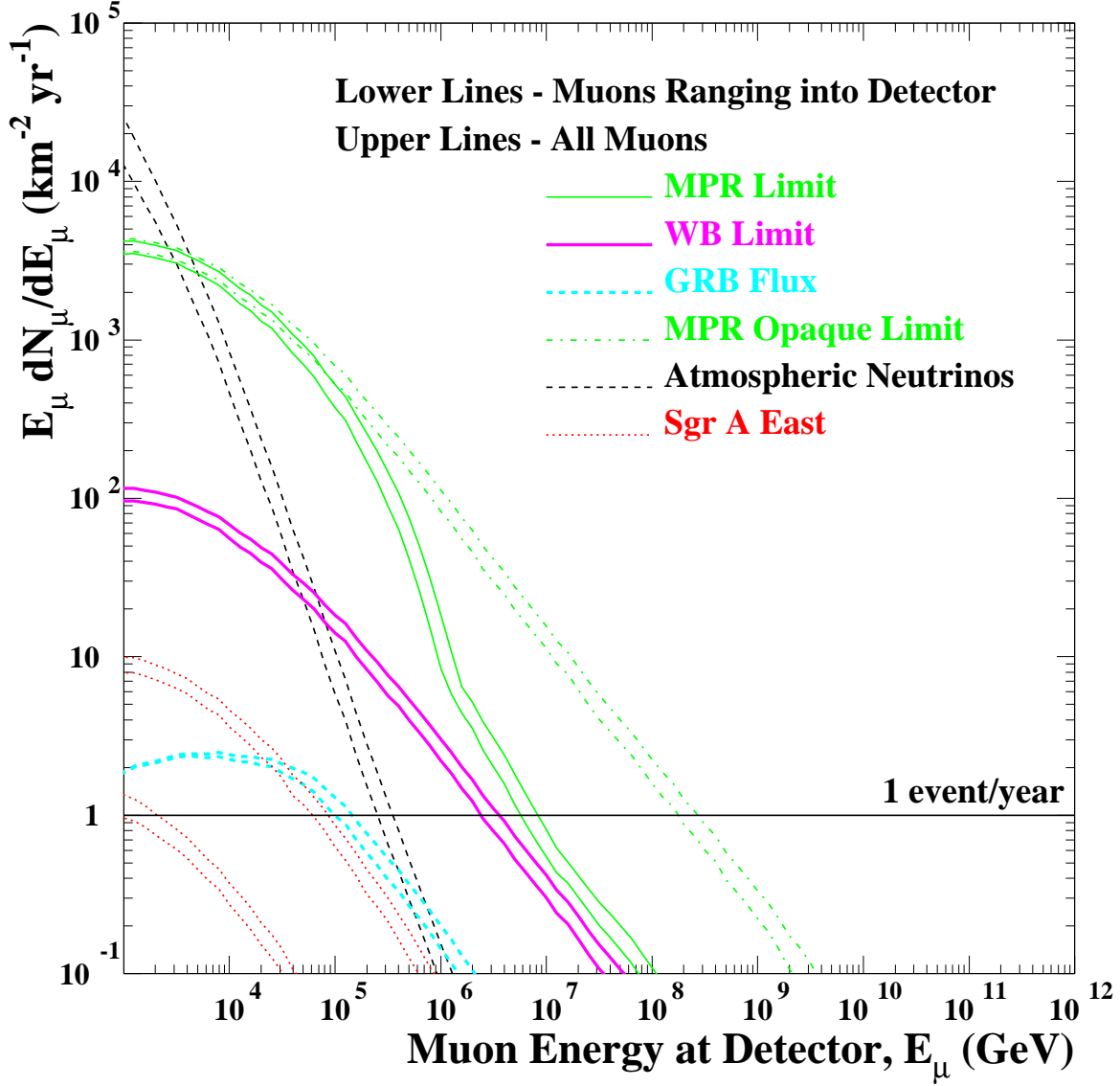}
\caption{Differential upgoing muon and anti-muon rate weighted by muon energy 
as a function of the muon energy when it enters the detector.
The measured energy of muons that range into the detector is less than
the initial energy.  Muons that start inside the detector make up about
30\% of the flux at high energy.}
\label{fig:muflux-detector}
\end{figure}

A stronger effect is noticed if we determine the flux as a function of 
the muon energy as it enters the detector. 
A very high-energy neutrino will make a very high-energy muon that travels
many kilometers losing energy as it goes.
If the track starts at random distance from the detector, the
measured energy will be distributed nearly uniformly 
between the initial energy and zero.
A high-energy muon will lose on average 1-1/e of its energy 
when traversing between 2.4 and 3 km of ice.
About each 2.7 km of ice traversed will move the muon down 
a natural logarithmic energy  interval of muon energy at the detector. 
We use the Lipari-Stanev Monte-Carlo \citep{lipari} to determine the 
muon range and the energy deposited in 100 m steps including fluctuations
from radiative processes.  
A random spot is chosen along the track to represent the point it enters
the detector.  
Figure~\ref{fig:muflux-detector} 
shows the differential muon plus anti-muon rate weighted by muon energy 
resulting from neutrino
interactions as a function of the muon energy as it enters the detector.

The atmospheric neutrino flux is a major background 
over most of the energy range
where events can be measured in a ${\rm km^2}$ per year.  
To detect the high-energy signals, good energy resolution is needed.
For point sources, the muon energy is used as a cut parameter to
optimize signal to background. 
For the diffuse flux, the resolution is needed 
to deconvolve the spectrum back to the neutrino energy.  
Where signals are not detected, 
a cut on muon energy is needed to set upper limits.

\subsection{Muon Energy Resolution}

\begin{figure}
\plotone{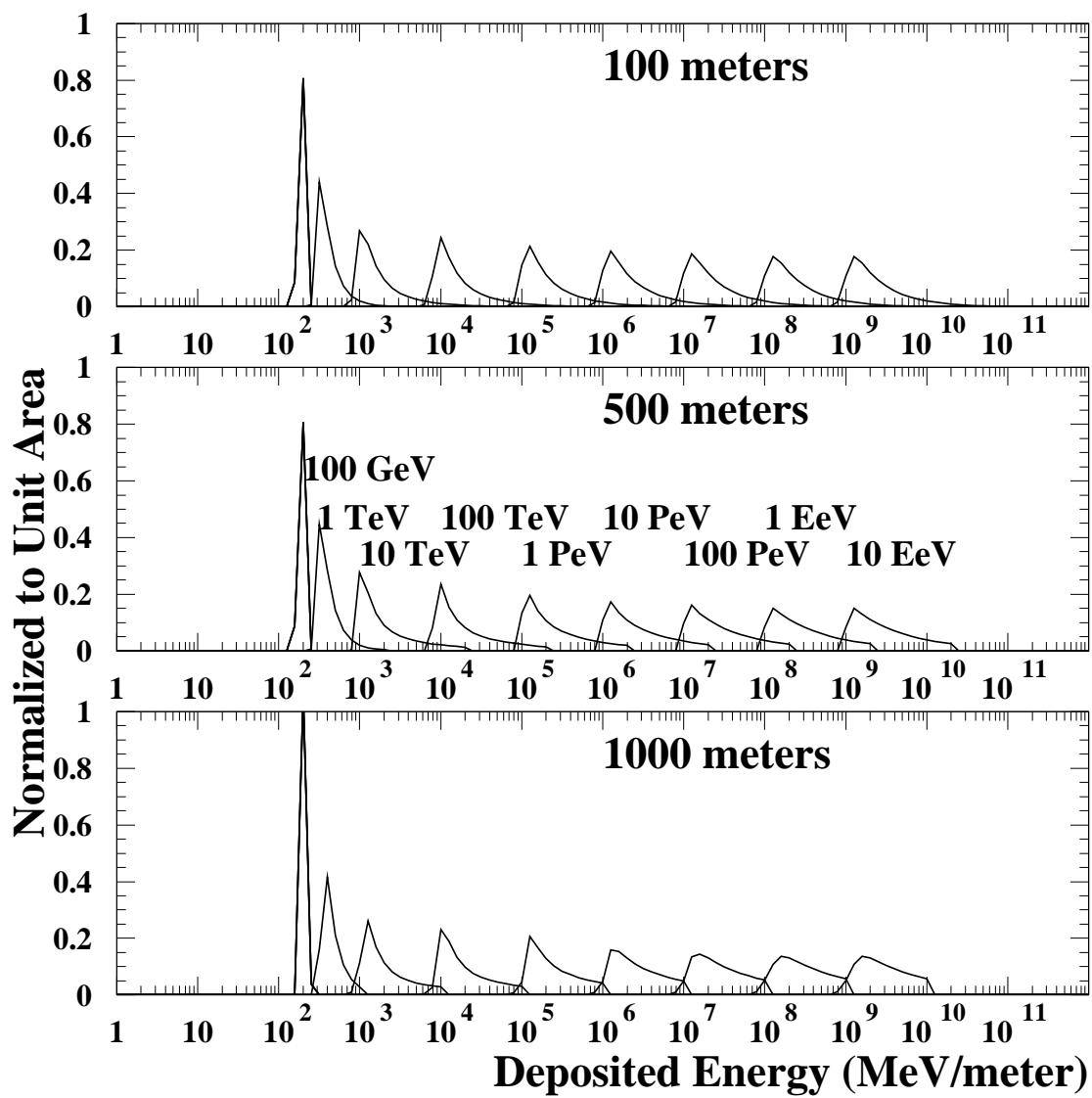}
\caption{Energy deposited in MeV/m for 
100m, 500m, and 1 km muon paths.  
Notice there is a strong dependence
in the deposited energy on the true muon energy until the muon energy
drops below about 100 GeV.}
\label{fig:ls}
\end{figure}

Muon energies below about 100 GeV are measured by their path length.
At higher energies, the path lengths become too long
and the amount of Cherenkov radiation from the track is used. 
The total Cherenkov radiation is nearly proportional to the
total energy deposition in the detector volume. 
Figure~\ref{fig:ls} shows the
deposited energy in MeV/m for 100, 500 and 1000 meter track lengths
at a variety of muon energies.   
The distribution is essentially a Gaussian with a long tail to larger energy 
deposition due to the stochastic nature of the energy loss processes\footnote{
The energy losses determined by the Lipari-Stanev Monte Carlo and more modern
codes \citep{mmc} are represented by
a delta function and a radiative tail.  Depending on the energy, there will
be a gap between these two parts (larger tracks have less gap).  We make
a smooth parameterization to avoid this gap.  We use the difference between 
the Lipari-Stanev standard prediction and our parameterizations 
as an estimate of the systematic uncertainty inherent in 
all currently available calculations.}.
The energy loss includes \citep{lipari} the usual ionization and knock-on 
electron processes important at lower energies and additional processes 
important at higher energies 
including pair production ($\propto \Delta E^{-\beta}$, $\beta$ varying from 1 to 3 as
$\Delta E$ increases) \cite{groommok},
bremsstrahlung ($\propto \Delta E^{-1}$), 
and photonuclear (roughly $\propto \Delta E^{-1.1}$),
all with a cut off at the total energy of the muon.
These processes are responsible for the energy dependent portion of $dE/dx$.
The numerous lower energy pairs (plus ionization) produce 
\footnote{The Central Limit Theorem states that 
the probability distribution of sum of variables drawn from probability
distributions with finite variances will tend towards the normal (Gaussian) 
distribution. 
The very numerous low-energy deposition processes will result in a nearly 
Gaussian peaked shape around the most probable energy loss for most energy 
depositions. 
The much rarer high-energy losses take many more samples to average down
and result in a long tail to higher-energy deposition. 
A true $1/\Delta E$ distribution would not tend to Gaussian as it has an 
infinite variance were it not for the maximum energy cut-off.}
the Gaussian-peaked shape for most energy losses. 
The bremsstrahlung and photonuclear processes and the rarer high-energy pairs 
produce a significant tail of much larger energy depositions.
If the most probable (or mean) energy deposition is used as a measure of 
the muon energy, a significant number of the much more abundant lower energy 
muons will be reconstructed as high energy muons. 
As a result a significant number of the abundant atmospheric muons 
would be reconstructed as much higher energy muons.
Longer sampling path lengths have a more truncated tail 
because there is a cut off where the muon is completely stopped.  
The deposited energy can not exceed the muon energy.  
A clear statistical relationship exists between detected and true muon energy.

\begin{figure}
\plotone{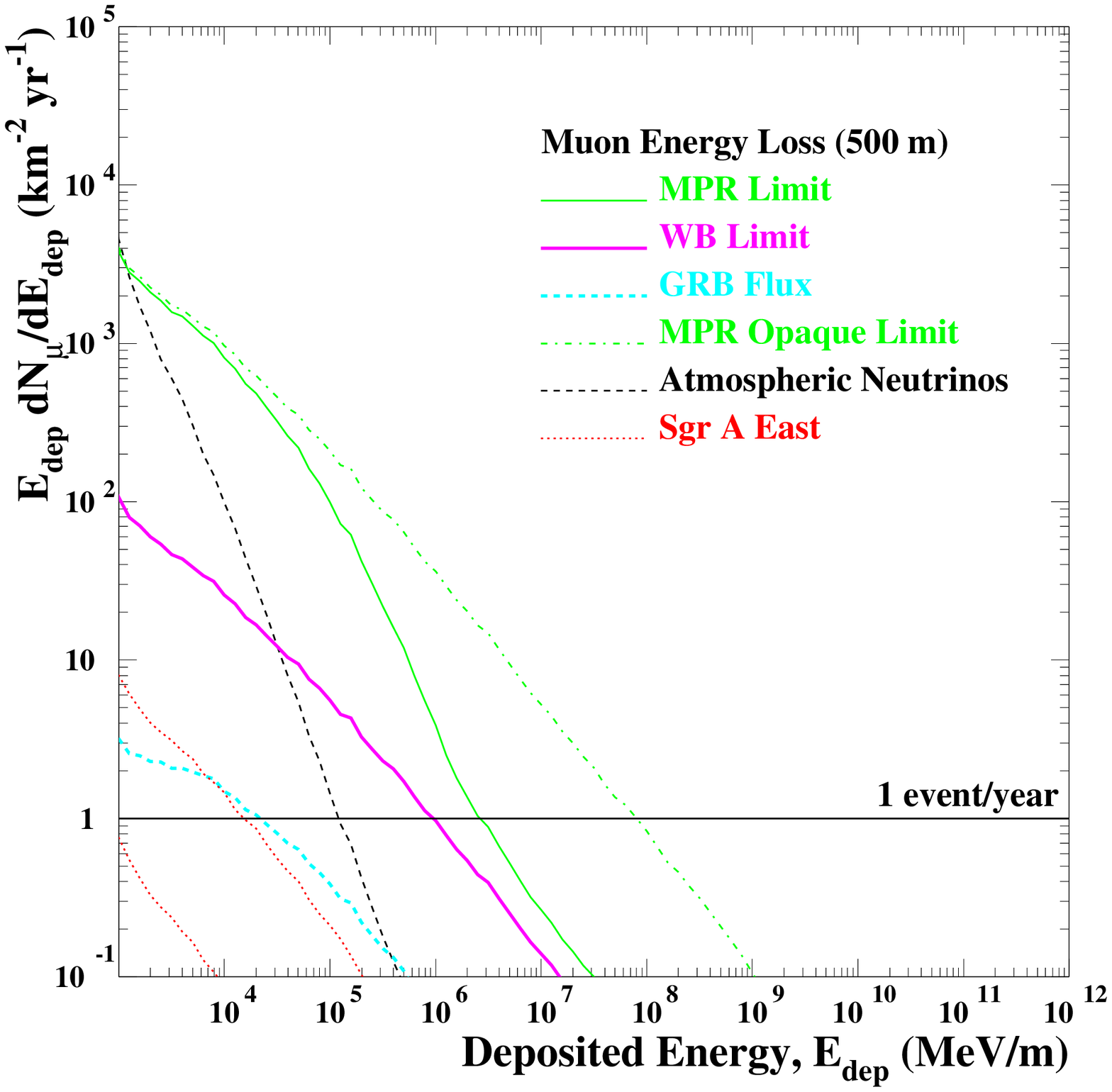}
\caption{Differential upgoing muon and anti-muon rate weighted by energy deposited in the 
detector as a function of the energy deposited in the detector.}
\label{fig:muflux-ls}
\end{figure}

An ideal detector could measure the deposited energy perfectly.  
If an ideal detector made many measurements with independent 100 meter
samples, then events that fluctuated early could be removed at the
expense of efficiency and the true muon energy could be measured  
using the samples up to the first large fluctuation.   
The Frejus detector \citep{frejus}
being a sampling calorimeter determined the muon energy based on fluctuations
in energy deposition. 
The large scale of more recent neutrino detectors makes independent 
sampling unrealistic.  
In an open geometry, such as Cherenkov detectors, the samples can not 
be cleanly separated with opaque barriers.  Long path lengths are an 
advantage because they minimize the sensitivity to the long tails in the
energy resolution.  Under sampling can reduce the advantage of long 
path lengths, however if the detector is only able to measure portions
of the energy deposition.
Figure \ref{fig:muflux-ls} 
shows the differential upgoing muon rate weighted by the energy deposited in the detector 
as a function of energy deposited in the
detector as radiation in a 500 meter track.

\begin{figure}
\plotone{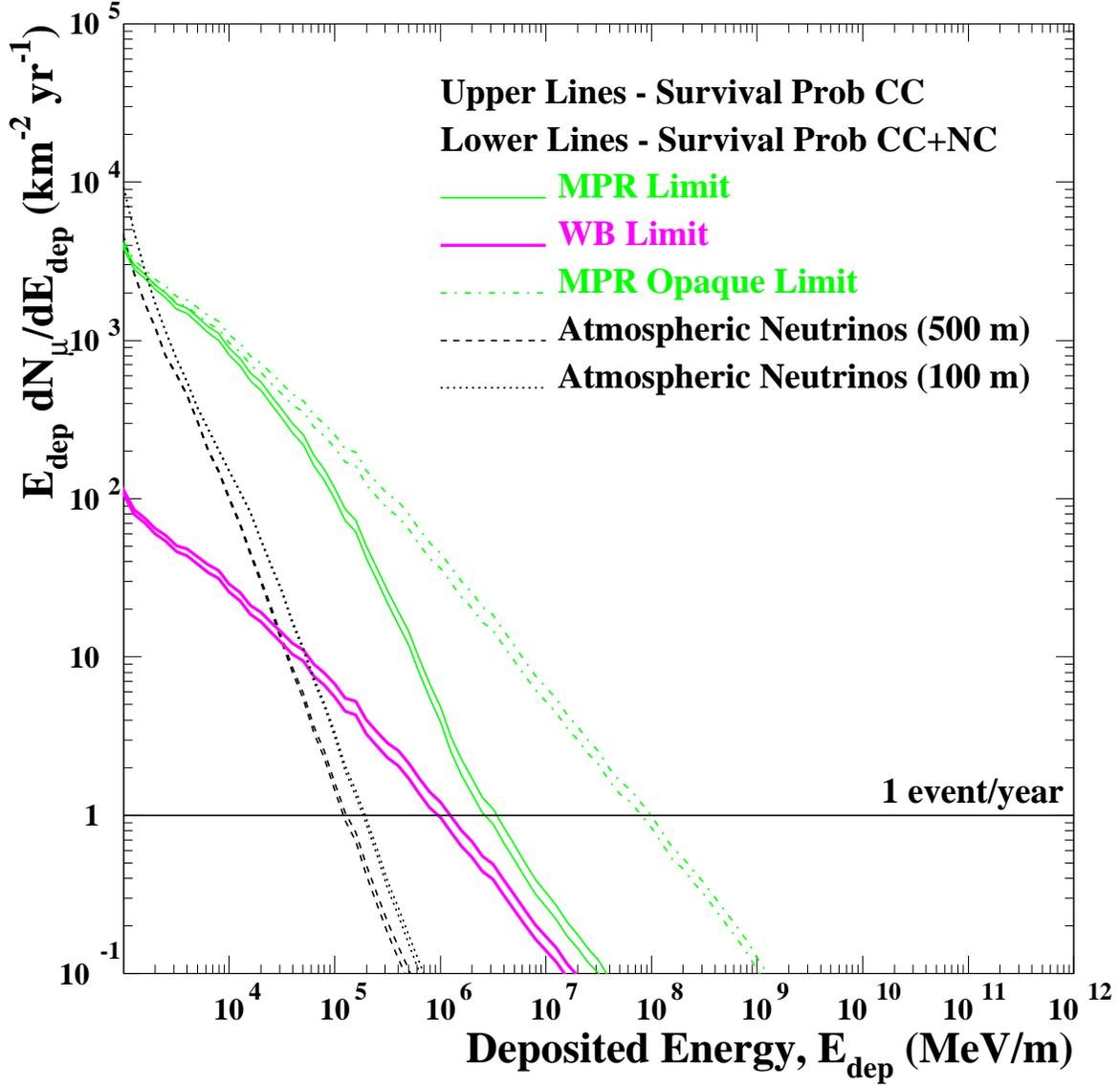}
\caption{The spectral dependence on survival probability and path
length.
Neutral currents (NC) primarily affect the overall normalization of the
hard spectra.
The spectrum dependence on path-length is shown for the steep atmospheric 
background where the resolution tails become important.  For 100 m paths
there is a kink in the distribution.  Without tails in the resolution,
such a change in slope would signal the discovery of a new source of 
neutrinos.  We use the
500 m resolution for our ideal detector and add the actual path length
in Section~\ref{sec:sens-real} for real detectors.}
\label{fig:muflux-ls-uncert}
\end{figure}

Figure~\ref{fig:muflux-ls-uncert} is included to show the main
systematic effects on the spectrum.
High energy muons are most sensitive to the NC effect on survival probability.
The shape of the atmospheric background depends more 
significantly on the length of the track.
In addition to the systematics shown on the plot, there is about 20\% 
theoretical 
uncertainty in the predicted atmospheric flux (see section~\ref{sec:atm}).

\subsection{Sensitivity of an Ideal km$^2$ Detector to Astrophysical Sources}
\label{sec:sens}

Event rates above a cut where signal and background are equal are listed
in Table~\ref{t:idealEvts}. 
Also included are rates for two other possible energy cuts: 
the energy where no background events are expected 
(atmospheric $<$ 0.1 event), and
the energy where 1 background event is expected.
Uncertainties in the rates come from several sources.  
There is a theoretical uncertainty on the atmospheric background of 
$\pm$20\%. The NC contribution to the survival probability is +10\%.  
Finally, the difference between Lipari-Stanev and a smooth parameterization
of the energy deposition
adds $\pm10\%$ for 500 meter tracks and $\pm50\%$ for 100 meter tracks.
The total uncertainties in the event rates quoted (500 m tracks) are: 
+14\%, -10\% for
signals and +24\%, -22\% for the atmospheric flux.

\begin{table}
\small
\begin{tabular}{|l||c|c|c|c|}
\hline
Muon Source  & events            & events          & energy (Mev/m)& events \\
             & E$>10^{5.5}$ MeV/m& E$>10^{4.95}$ MeV/m & S/B=1 & E$>$S/B=1 \\
\hline\hline
Atmospheric Neutrinos    & 0.11  & 1     & & \\ 
WB Opt Thin Limit        & 3.0   & 7.8   & $10^{4.3}$ & 23.8   \\
MPR Opt Thin Limit       & 15.5  & 90.0  & $10^{2.8}$ & 7500.0 \\
MPR Opaque Limit         & 112.0 & 300.0 & $10^{2.9}$ & 7050.0 \\
MPR Source ($10^3 GeV)$     & 10$^{-12}$ & 10$^{-7}$ &   & \\
MPR Source ($10^4 GeV)$     & 0.0003  & 0.31& $10^{3.0}$ & 2500.0 \\
MPR Source ($10^5 GeV)$     & 3.4     & 60.0& $10^{2.9}$ & 4000.0 \\
MPR Source ($10^6 GeV)$     & 5.8   & 21.0  & $10^{3.9}$ & 76.0   \\
MPR Source ($10^7 GeV)$     & 2.5   & 4.5   & $10^{4.5}$ & 6.4   \\
MPR Source ($10^8 GeV)$     & 0.9   & 1.3   & $10^{4.9}$ & 1.4   \\
MPR Source ($10^{11} GeV)$    & 0.1   & 0.1   & &   \\
\hline
\end{tabular}
\caption{Event rates for 500 meter long muon tracks in an ideal 
km$^2$ detector per year. 
The background is composed of atmospheric neutrinos.  
Background-free limits can be estimated for deposited 
energy $ > 10^{5.5}$ MeV/m.
A single background event is expected for energy $>10^{4.95}$ MeV/m.  
Finally, the energy and rates where the signal equals background (S/B=1) 
are listed.
Uncertainties in event rates are +24\%, -22\% for atmospheric background and 
+14\%, -10\% for the other signals.}
\label{t:idealEvts}
\end{table}

\begin{figure}
\plotone{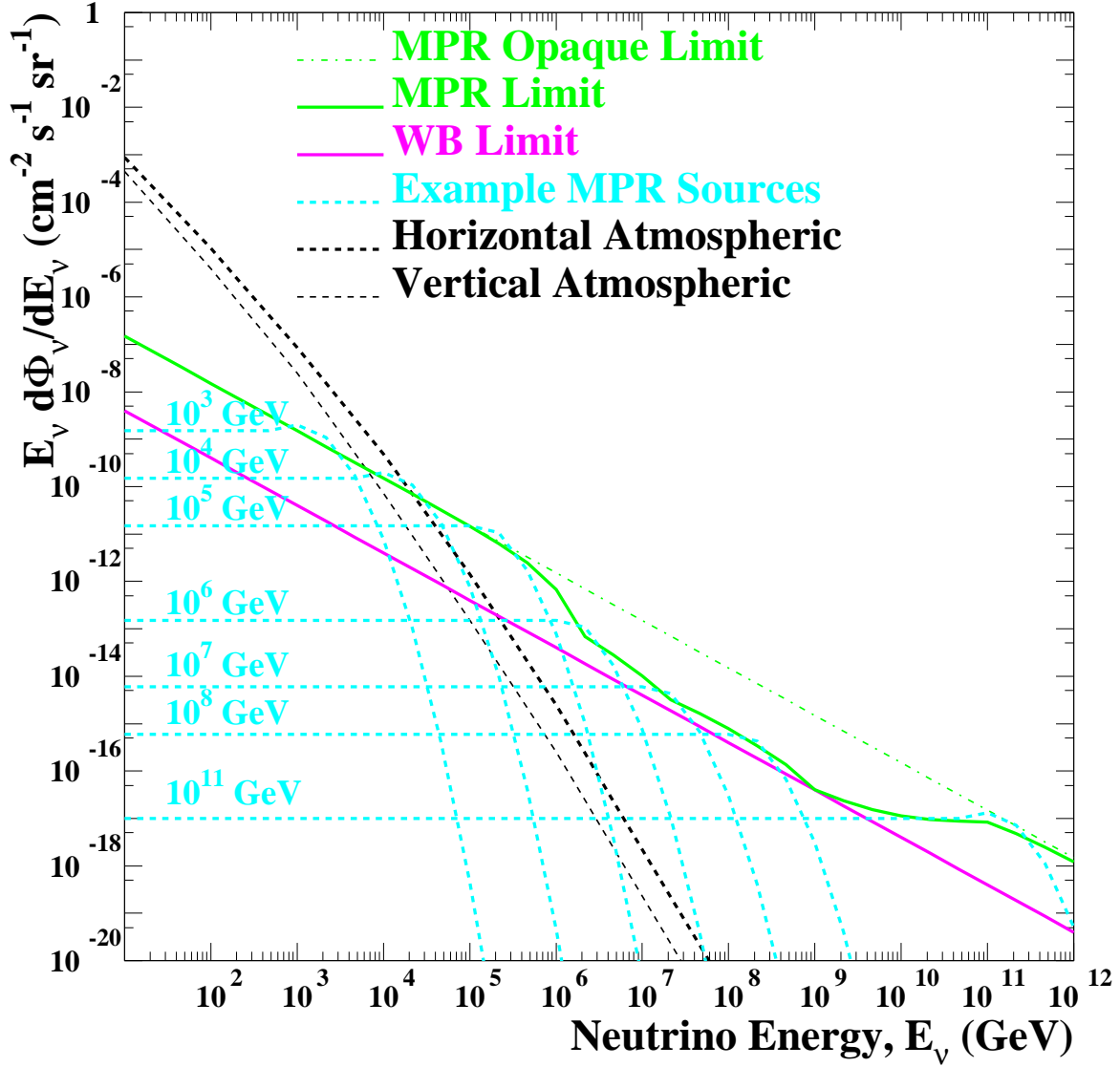}
\caption{Possible sources making up the MPR flux.  Sources are E$^{-1}$ spectra
cut-off by a gaussian of width 0.3.    
Amplitudes GeV/(cm$^2$ s sr) (cut-off energy(GeV)) $1.5 \times 10^{-6}$ (10$^{3}$), 
$1.5 \times 10^{-6}$ (10$^{4}$), $1.5 \times 10^{-6}$ (10$^{5}$), 
$1.5 \times 10^{-7}$ (10$^{6}$), $6.0 \times 10^{-8}$ (10$^{7}$), 
$6.0 \times 10^{-8}$ (10$^{8}$), and $1.0 \times 10^{-6}$ (10$^{11}$).  }
\label{fig:nuflux-sources}
\end{figure}

\begin{figure}
\plotone{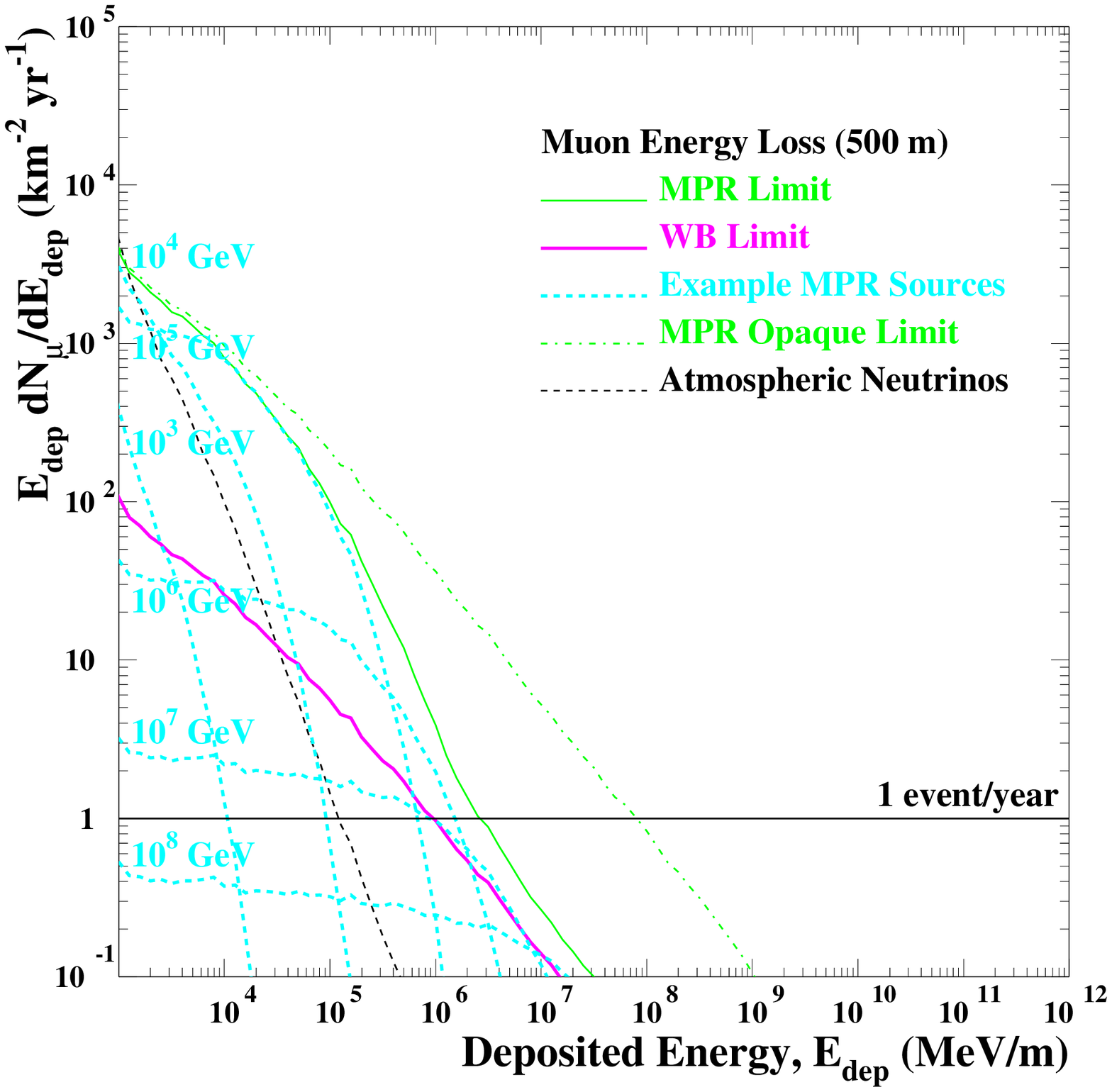}
\caption{Differential upgoing muon rate weighted by deposited energy as a function 
of energy deposited in the detector for
a series of example sources that could make up the MPR flux.
}
\label{fig:muflux-sources}
\end{figure}

Also included in the table are a series of sources consistent with the
MPR optically thin limit.  Since this limit is not a power law, it is
a composite of sources at each energy. The maximum
power-law spectrum consistent at all energies is the WB limit.
Figure~\ref{fig:nuflux-sources} shows the neutrino flux for 
these made-up sources.  We generously
give them a E$^{-1}$ isotropic spectrum, cut-off by a gaussian shape with means varying 
from $10^{3.0}$ to $10^{11.5}$ GeV, and widths of $\sigma = 0.3$.  
Their normalization varies between the MPR thin limit and WB limit.
Figure~\ref{fig:muflux-sources} shows the differential upgoing muon rate
weighted by the energy deposited in the detector for these sources
as a function of deposited energy in the detector.  From the numbers tabulated
in Table~\ref{t:idealEvts} it is clear that the a km$^2$ 
detector will be sensitive to neutrino fluxes with incident energies 
between $10^{5}$ and $10^{7}$ GeV.  Higher fluxes have been ruled out by 
the cosmic ray spectrum and lower fluxes are buried below the atmospheric
background.  

\begin{sloppypar}
In the absence of signal events, flux
limits can be set \cite{pdg2, FeldmanCousins} based
on the number of events observed, $n_{obs}$, by an experiment and 
knowledge of the mean background expected, $b$.  The poisson probability
that an observation is consistent with a mean signal, $s$ is given by
\begin{equation}
P(n_{obs}|s) = (s + b)^{n_{obs}}exp[-(s + b)]/n_{obs}!
\end{equation}
\end{sloppypar}

Integrating this over all signals up to a confidence level, $CL$, gives
the standard confidence belt, $s_{CL}(n_{obs},b)$.  
Feldman and Cousins suggest that this belt 
be modified to avoid flip-flopping between one and two-sided intervals
based on the experiment performed.  Here we choose the simpler approach
of always using a one-sided limit.  The Feldman and Cousins approach
leads to $\sim 10$\% weaker limits for very low statistics experiments.
The experimental flux limit is defined as
\begin{equation}
\Phi_{limit} = \phi_{s0} \times \frac{s_{CL}(n_{obs},b)}{s}.
\end{equation}
where $\phi_{s0}$ is the theoretical neutrino signal flux.
Before the experiment is performed, there are no observed events.
It is still interesting to 
determine the average $<s_{CL}(n_{obs},b)>$ of a
collection of proposed experiments having only background events.
\citep{hill}
\begin{equation}
<s_{CL}(n_{obs},b)> = \sum_{n_{obs}=0}^{\infty} 
s_{CL}(n_{obs},b) P(n_{obs}|b) 
\end{equation}
We define the average limit that a detector can place on signal fluxes 
to be the sensitivity of the detector, $\Phi_{sensitivity}$.  
\begin{equation}
\Phi_{sensitivity} = \phi_{s0} \times \frac{<s_{CL}(n_{obs},b)>}{s}.
\end{equation}
The ratio $\phi_{s0}/s$ is the detection transfer function.  
The magnitude of the signal flux cancels in the ratio leaving only the
sensitivity to the spectral shape.  

\subsubsection{Diffuse Limits}
\begin{sloppypar}
Table~\ref{t:idealSens} shows the sensitivity (95\% CL) of
an ideal detector of km$^2$ incident area to neutrino fluxes
and various spectral shapes as suggested in \citep{mpr}.
Included in these estimates are systematic 
uncertainties in the background normalization.
\end{sloppypar}

Systematic uncertainties affecting the background and 
signal bound our knowledge of the measurement.
For example, an experiment that expects 1000 background events can
easily discover a signal with 100 events, even with poisson sampling
statistics.  But this same experiment with a 10\% uncertainty in the
background is barely sensitive to a signal with 200 events.
The limits quoted in Table~\ref{t:idealSens} include Poisson statistics
as described above, but limit the sensitive region to where the signal is
larger than the 2 $\sigma$ uncertainty on the background.  

\begin{table}
\small
\begin{tabular}{|l||c|c|c|c|c|}
\hline
Muon Source

Flux         & Optimized & Atmospheric& Signal & Original  & km$^2$ Sensitivity \\
             & Energy Cut& Background, $b$& $s$& Amplitude & GeV/cm$^2$ s sr\\
	     & (MeV/m)   &   (events) & (events) & GeV/cm$^2$ s sr &  (95\% CL)      \\
\hline\hline
WB Opt Thin Limit  & $10^{4.3}$ & 16.1 & 23.8 & $4.0 \times 10^{-8} E^-2$   & $1.3 \times 10^{-8}$ \\
MPR Opt Thin Limit & $10^{4.3}$ & 16.1 & 487. & varies                 & $\times0.016$ \\
MPR Opaque Limit   & $10^{4.3}$ & 16.1 & 893. & $1.5 \times 10^{-6} E^-2$   & $1.3 \times 10^{-8}$   \\
MPR Source (10$^{4}$ GeV)&$10^{4.3}$ & 16.1 & 36.4 & $1.5 \times 10^{-6}$ & $2.1 \times 10^{-7}$ \\
MPR Source (10$^{5}$ GeV)&$10^{4.3}$ & 16.1 & 448. & $1.5 \times 10^{-6}$ & $2.6 \times 10^{-8}$ \\
MPR Source (10$^{6}$ GeV)&$10^{4.6}$ & 4.47 & 36.1 & $1.5 \times 10^{-7}$ & $2.2 \times 10^{-8}$ \\
MPR Source (10$^{7}$ GeV)&$10^{5.0}$ & 0.83 & 4.31 & $6.0 \times 10^{-8}$ & $4.4 \times 10^{-8}$ \\
MPR Source (10$^{8}$ GeV)&$10^{5.3}$ & 0.24 & 1.07 & $6.0 \times 10^{-8}$ & $1.4 \times 10^{-7}$ \\
\hline
\end{tabular}
\caption{Sensitivity of a km$^2$ detector to fluxes with a variety of
spectral shapes based on 1 year statistics.}
\label{t:idealSens}
\end{table}

We find that a detector with km$^2$ incident area will be sensitive to a spectral
flux three times smaller than the WB limit.  From the MPR sources, we can see
that the limit is most sensitive to neutrinos of ~1 PeV.  Neutrinos 
between $10^{5}$ and $10^{7}$ contribute most to the signal rate.

\subsubsection{Point Source Detection - GRB}

One easy way to reduce the background in these experiments is to narrow
the search bin from half the sky to the characteristic size of the detected
angular resolution.  The intrinsic resolution of a muon's direction with
respect to the neutrino is about 1 square degree.  In this way, we
can divide half the sky into 20,628 one degree square patches of sky.
There are two kinds of searches.  The first involves looking for 
neutrinos from sources that are known to exist.  The second involves looking
for sources anywhere on the sky.

The GRB flux is a case where both the time and location of the burst is
known {\it apriori}.  In this case, we take all known bursts and search in one
degree bins, coincident in time.  The integrated GRB signal from 
Figure~\ref{fig:muflux-ls} is 15 events per year.  
The number of GRB detected in
a year depends on the sensitivity of experiments like BATSE or MILAGRO.
Based on expectations of these detectors, we estimate somewhere in the
range of $10^2-10^3$ GRBs per year.  Assuming that all 15 muons above were
produced in some fraction of these GRBs, we find that the expected background
in all GRB events is 0.015 muons.  A $5\sigma$ discovery can be made even
if the GRB flux is reduced by a factor of 5.  

Somewhat surprising is the robustness of this result to variations in 
the number of detected GRBs and the time-scale of the event.  Depending
on the number of GRBs detected by other experiments, the background can 
change by an order of magnitude. 
Similarly, if the neutrinos arrive over a 24 hour period instead of a 1 second
pulse (used in the above calculation -- see section~\ref{sec:grb}) then the 
background is 86000 times larger.
In this case, the best limit comes from applying an energy cut at 
$10^{2.4}$ MeV/m which leaves only 0.012 background events and 13 GRB
neutrino induced muons.  If this energy cut is not possible, then the
background is too large, and there is
no way to find GRB muons that arrive on a long time scale.

\subsubsection{Point Source Detection - Sgr A East}

Sgr A East is an example of a bright galactic source.  It is known
to vary in brightness, but on a time-scale of months so that 
it is reasonable to expect measurements to
integrate the neutrino signal over a full year.  
We estimate that the Sgr A East rate
is between 0 and 40 neutrino induced muons per year
in an ideal km$^2$ detector at a latitude of +35 degrees.  
The Atmospheric Neutrino 
background is reduced by knowing the location of the source.  
The best $5\sigma$ limit comes from applying an energy cut at 
$10^{2.4}$ MeV/m which leaves only 4 background events and 33
neutrino induced muons from Sgr A East with the hardest hypothesized
proton spectrum.  Such a detector (equivalent to a km scale Mediterranean-based
detector) will be sensitive to a source even
half as bright.  The softest spectrum is about ten times dimmer and 
would not be detectable.

\subsubsection{Point Source Detection - AGN}

Another type of point-source search involves looking for sources without 
prior knowledge of location or time.  Here we reduce the up-going 
atmospheric neutrino background by 20,628 search bins on the sky. 
The signal is also divided among an unknown number of sources.

There is an art to choosing bins on the sky.  If the bins are chosen
before the experiment is performed, then sources will not fall in just
one bin, and the search is not efficient.  If sliding windows are
used to find spots with the largest number of events, then the search
is biased to the locations where the background has clustered.  A poor-man's
alternative is to consider a fixed array of search bins, but to perform
100 searches with each shifted by 1/10th of a degree in azimuth or zenith
from the previous search.  This is a close approximation to 100 independent
searches of the sky.  The signal will be 90\% contained in at least one 
search, and can therefore be approximated by the true signal, ignoring the
cases where the signal is partially contained in a different search.  
The effect is that instead of performing 20,628 experiments for each
spot on the sky, we perform 2,062,800 experiments.  To avoid mistaking a
background fluctuation for signal, we calculate $6\sigma$ limits which
will only be wrong one time out of 5 million.

If the entire isotropic diffuse flux is produced entirely from one point source
(within a single degree-square bin) then
an ideal km$^2$ detector will be sensitive to a flux 
$1.4\times10^{-9}$E$^{-2}$ GeV/(cm$^2$ sr s) at $6\sigma$.
This is 28 times lower than the WB limit.  Since the rate of 
neutrinos is expected to be about 5 times lower than the limit,
we conclude that this rate is expected and easily detected.
If, however,
the flux is divided between 10 bright sources, then we will only be
sensitive to a flux 2.8 times lower than WB, and a point source discovery
would indicate new physics.   This limit scales linearly with the
number of sources that contribute to the flux; so, for example,
$10^3$ sources are the maximum that may be detected by an ideal km$^2$
detector because more would violate the MPR Opaque Limit.  

The diffuse limits have been treated as isotropic.  From 
Figure~\ref{fig:survival}, one can see that the diffuse flux
($10^5-10^7$ GeV) is biased toward the horizon.  
Astronomy with neutrinos relies on the flux being divided among a 
handful of bright point sources located within a few tens of degrees
of the horizon.

\section{Sensitivity of Proposed and Existing Detectors}
\label{sec:sens-real}

The above calculations used an ideal geometry of a detector with km$^2$
incident area and 500 meters long for all zenith angles.  We now include the
geometry of proposed detectors to determine their acceptance.  Combining
the irreducible physics effects with the detector acceptance allows us
to determine the best possible limit for such detectors.

Not included
are the effects of specific detector designs which can only degrade the 
sensitivities.  
In real life detectors
tend to be cylinders or spheres sunk into deep water, ice or caves.  
The interaction probability of the rock below or surrounding the 
detectors and the passive detecting medium of the instrument have to
be considered.  Particularly important effects which are not addressed in
this paper are the number of sampling elements, 
the sensitivity of the detecting elements, the uniformity
of the detecting medium, and the conversion of ``deposited radiation'' into 
a measurable light spectrum.   

\begin{figure}
\plotone{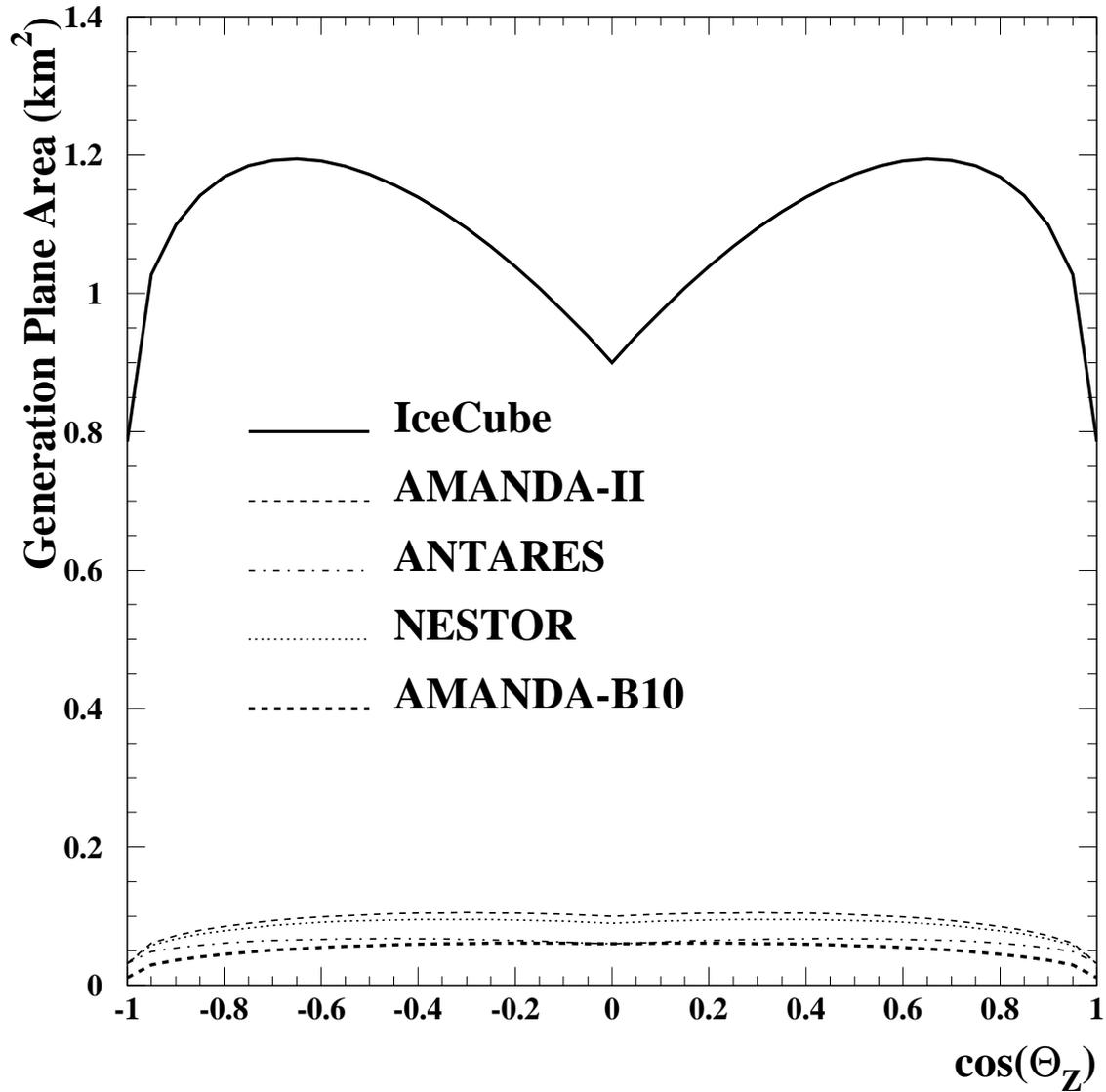}
\caption{Geometrical dependence of the IceCube, ANTARES, NESTOR and  
AMANDA detectors as a function of zenith angle.
Dimensions are for a cylinder of (depth, height, diameter) in meters: 
IceCube (1850, 900, 1000), 
AMANDA-II (1740, 500, 200), 
ANTARES (2250, 300, 200), 
NESTOR (3400, 450, 200), 
AMANDA-B10 (1740, 500, 120).}
\label{fig:geo1}
\end{figure}

\begin{figure}
\plotone{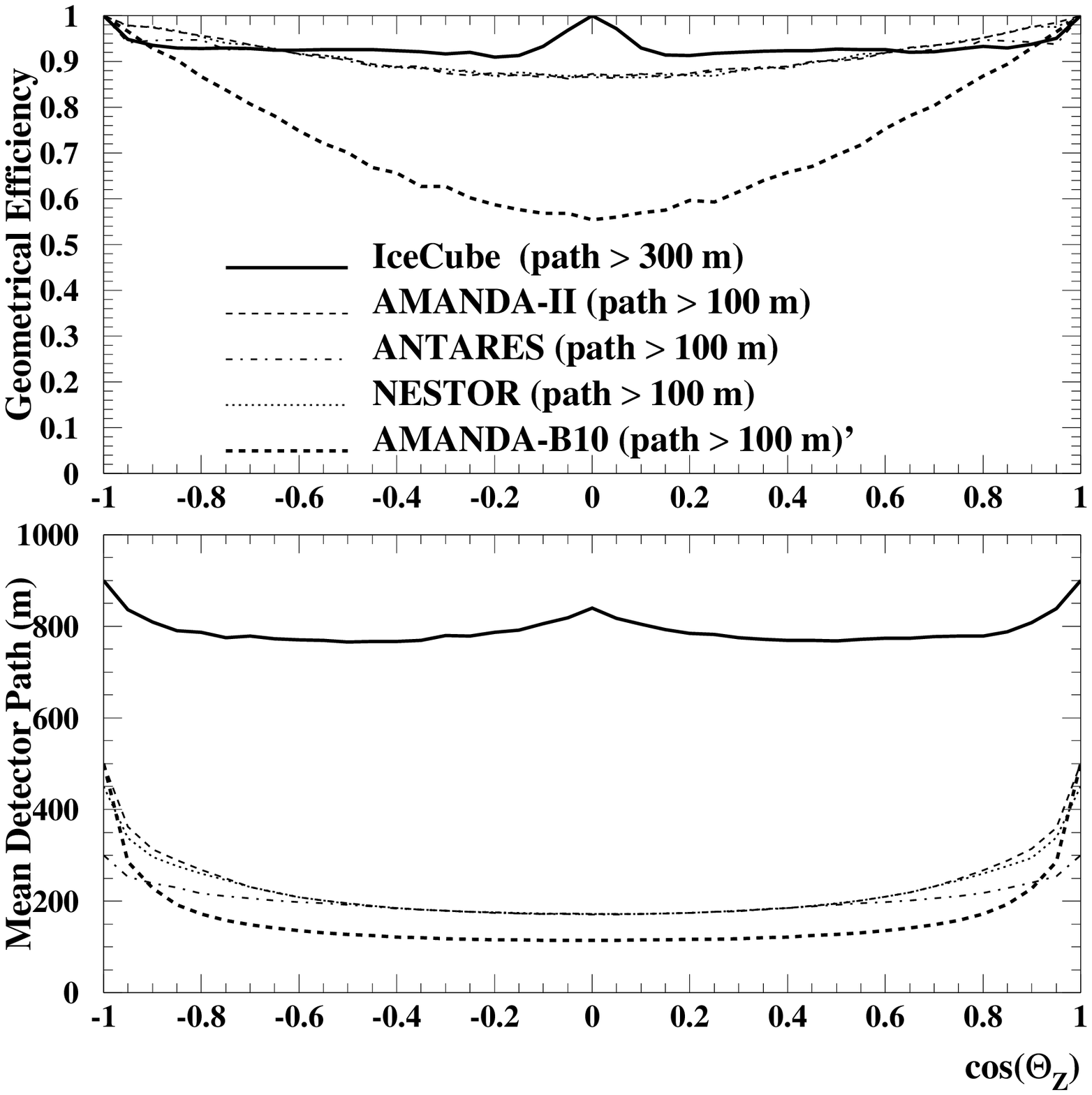}
\caption{Geometrical efficiency for tracks longer than either 100
or 300 meters as a function of zenith angle.}
\label{fig:geo2}
\end{figure}

Figure~\ref{fig:geo1} shows the geometrical profile of 
IceCube, AMANDA-II, ANTARES, NESTOR, and AMANDA-B10
as a function of zenith angle.  Figure ~\ref{fig:geo2} shows the 
mean detector path length and efficiency assuming a reasonable 
minimum path length.
IceCube is by far the largest, with essentially km$^2$ acceptance.
AMANDA-II, ANTARES, and NESTOR are all of similar size and
and AMANDA-B10 is the smallest.

Table~\ref{t:realSens} lists the sensitivities of these detectors to 
an E$^{-2}$ flux of high-energy neutrinos.
These estimates include the effects of neutrino attenuation in the earth,
muon transport, and fluctuations in energy deposition.  For IceCube, a 300 m
minimum track is required to reduce the systematic effects of the long tails
in the resolution.  For the others, a 100 m minimum track is required.  For
the smaller detectors, the irreducible systematic uncertainty is quite
important.  For IceCube, we use +24\% for the atmospheric background systematic
uncertainty, and for the others we use +50\% (see section~\ref{sec:sens}).
Limits without systematics are 10\% better for IceCube, and 30\% better
for the smaller detectors.

\begin{table}
\small
\begin{tabular}{|l||c|c|c|c|c|c|c|c|}
\hline
             & \multicolumn{2}{c|}{Optimized}   & 
               \multicolumn{2}{c|}{Atmospheric} & 
               \multicolumn{2}{c|}{WB Signal} & 
               \multicolumn{2}{c|}{Sensitivity} \\ 
Detector     & \multicolumn{2}{c|}{Energy Cut} & 
               \multicolumn{2}{c|}{Background, $b$} & 
               \multicolumn{2}{c|}{$s$} &
               \multicolumn{2}{c|}{ }   \\
	     & \multicolumn{2}{c|}{(MeV/m)}    &   
               \multicolumn{2}{c|}{(events)} &  
               \multicolumn{2}{c|}{(events)} &  
               \multicolumn{2}{c|}{(GeV/cm$^2$ s sr)} \\
\hline \multicolumn{1}{|r||}{CL}
             & $90\%$ & $95\%$ & $90\%$ & $95\%$ & $90\%$  
& $95\%$ & $90\%$ & $95\%$ \\
\hline\hline
IceCube     & $10^{4.3}$ & $10^{4.2}$ & 9.1 & 13.9 & 19.1 & 22.7 &
$1.1 \times 10^{-8}$E$^{-2}$ & $1.4 \times 10^{-8}$E$^{-2}$ \\ 
AMANDA-II   & $10^{4.1}$ & $10^{4.0}$ & 2.9 & 4.3 & 2.2 & 2.6 &
$6.5 \times 10^{-8}$E$^{-2}$ & $8.0 \times 10^{-8}$E$^{-2}$ \\ 
ANTARES     & $10^{4.0}$ & $10^{3.9}$ & 2.8 & 4.1 & 1.6 & 1.9 &
$8.7 \times 10^{-8}$E$^{-2}$ & $1.1 \times 10^{-7}$E$^{-2}$ \\ 
NESTOR      & $10^{4.1}$ & $10^{3.9}$ & 2.7 & 5.8 & 2.0 & 2.8 &
$7.0 \times 10^{-8}$E$^{-2}$ & $8.2 \times 10^{-8}$E$^{-2}$ \\ 
AMANDA-B10  & $10^{3.9}$ & $10^{3.7}$ & 2.8 & 5.4 & 1.2 & 1.7 &
$1.2 \times 10^{-7}$E$^{-2}$ & $1.3 \times 10^{-7}$E$^{-2}$ \\ 
\hline
\end{tabular}
\caption{Sensitivity of existing and proposed detectors to fluxes with 
E$_{\nu}^{-2}$ spectral shapes based on 1 year statistics.}
\label{t:realSens}
\end{table}

  We state in section~\ref{sec:summary}
that to ensure the discovery of neutrinos from
astrophysical sources one needs a detector sensitive to about
one fifth of the WB flux.  We find that current and future detectors 
are at most sensitive to one third of the WB flux.
IceCube will reach a sensitivity of one fifth the WB flux after 2-3
years of 100\% efficient operation. 
To characterize the source luminosity or energetics 
would require additional factors of 10-100 in rate.

\section{Conclusions}
\begin{sloppypar}
We have summarized the muon neutrino plus muon anti-neutrino fluxes from 
astrophysical sources. 
We include fluxes predicted by models within the 
particle physics standard model
and do not include the ones predicted by exotic models. In this way the
muon neutrino flux is constrained by the Waxman and Bahcall limit  (WB) \citep{WB1}
for energies above $10^6$ GeV. The ``thin source'' (see \ref{sec:mpr}) limit
from Mannheim, Protheroe and Rachen \citep{mpr} can slightly loosen this limit
for energies between $10^6$ and $10^7$ GeV and above $10^9$ GeV. The ``thick
source'' limit from these authors \citep{mpr} is shown to take
into consideration sources that are unlikely to exist if they behave
as expected by standard model physics.
Any sources violating the thin limit must be modeled with physics
beyond the standard model.  These are in the
exotic sources category and will be considered in a future analysis.
\end{sloppypar}

The neutrino signal is given by secondary muons produced in a charged
current interaction of the neutrino with either the rock below the detector or
the ice or water inside or surrounding the detector. 
We translate the muon neutrino
event rate to a muon event rate and show our results in 
Figure~\ref{fig:muflux-ls}. 

\begin{sloppypar}
From these rates we determine the sensitivity
(see tables~\ref{t:idealSens}~and~\ref{t:realSens}) for an ideal 
detector of Km$^2$ incident area as well as for current and proposed 
experiments (AMANDA-B10, AMANDA-II, NESTOR, ANTARES and IceCube).

Among the current experiments, AMANDA-B10 is the only one with
a reported limit.  At the 90\% CL they find a limit of
$0.9\times10^{-6}$E$^{-2}$ GeV/(cm$^2$ sr s) \citep{hill}.
This limit is based on 137 days of live-time during 1997.  For comparison,
we find $2.7\times10^{-7}$E$^{-2}$ GeV/(cm$^2$ sr s) at 90\%
CL for the same live-time statistics.
This is consistent with their result if one considers that the
instrument has additional resolution effects to be taken into
consideration.  The sensitivities listed in Table~\ref{t:realSens}
are for 1 year of live-time and are more than 2 times lower.

The predicted sensitivity for Amanda-II is
$7\times 10^{-8}$E$^{-2}$ GeV/(cm$^2$ sr s) for 2 years of operation
\citep{barwick}.  Assuming that two years would produce a better limit
than one year, we conclude that the instrument has additional
resolution issues.
ANTARES predicts a sensitivity of
$2\times 10^{-7}$E$^{-2}$ GeV/(cm$^2$ sr s) at 99.99\% CL (5$\sigma$)
\citep{antares}. This is in agreement with our estimate
which includes only irreducible physics and gross geometrical effects.
The predicted sensitivity for IceCube at 90\% CL is
$9.5\times 10^{-9}$E$^{-2}$ GeV/(cm$^2$ sr s) \citep{leuthold}
which agrees with our estimate if we calculate without systematic
uncertainties. Since the Leuthold estimate includes a full
detector simulation, we conclude that the detector comes close to the
irreducible physical limit.
It is important to note that these predicted sensitivities are
based on 100\% duty cycles and do not include
the dead time due to trigger, maintenance and other normal experimental
procedures.
We point out that our estimates are optimistic since we do not
include additional degradation due to instrumental effects as
described in the previous section.
\end{sloppypar}

  The most promising flux to be measured is that from GRB neutrinos.
The background in point-source searches is greatly reduced by spatial
and temporal localization.  Discovery of GRB neutrinos at the 5$\sigma$
level is predicted to be possible in an ideal Km$^2$ detector according
to current flux models.  
Predicted neutrino fluxes for Sgr A East vary by an order of magnitude. 
The brightest estimates from Sgr A East also 
yield robust detections in a similar detector at Mediterranean
latitudes.
However, if there is no prior knowledge of location and time, detection
of point sources relies on the flux being divided among no more than a handful of
bright sources.

Discovery of neutrinos from high energy astrophysical sources, ie, the ones
which are able to produce particles with energies around $10^{19}$ eV,
will likely require a detector designed to a
sensitivity of one 
fifth of the WB limit. This limit is at least five times conservative (see
Section~\ref{sec:WB}) and such a sensitivity would provide possibility 
of detection.
From all detectors analyzed, IceCube comes closest
to this sensitivity, being able to measure a flux 3 times lower than 
the WB limit in one year and 5 times lower after 2-3 years of full operation.

\acknowledgements
We thank Christopher Spitzer for his support in the numerical analysis, 
Dmitry Chirkin for his insight into muon radiation and Steve Barwick,
Willi Chinowsky, Azriel Goldschmidt and Jozsef Ludvig for useful comments.
We also thank NERSC for supporting the calculations in
this paper with high-performance linux computing.

This work supported by NSF Grants KDI 9872979 and Physics/Polar Programs 
0071886
and in part by the Director, Office of
Energy Research, Office of High Energy and Nuclear Physics, Division of
High Energy Physics of the U.S. Department of Energy under Contract No.
DE-AC03-76SF00098 through the Lawrence Berkeley National Laboratory.

\end{document}